\def\footnoterule{\kern -3pt \hrule width 0truein \kern 2.4pt}
\font\twelvermb=cmbx10 at 12pt \font\eightrmb=cmbx10 at 8pt

\bigskip
\bigskip
\noindent{\twelvermb Calculation of some determinants using the $s$-shifted factorial}\hfill

\bigskip
\bigskip
{\leftskip=1.5cm
\noindent {\bf Jean-Marie Normand}

\medskip
{\eightrmb
\noindent Service de Physique Th\'eorique, CEA/DSM/SPhT - CNRS/SPM/URA 2306

\noindent CEA/Saclay, F-91191 Gif-sur-Yvette Cedex, France

\medskip
\noindent E-mail: jean-marie.normand@cea.fr
}

\bigskip
\bigskip
\noindent{\bf Abstract}

\noindent Several determinants with gamma functions as elements are evaluated.
These kinds of determinants are encountered, for example, in the computation of the probability density of
the determinant of random matrices.
The $s$-shifted factorial is defined as a generalization
for non-negative integers of the power function, the rising factorial (or Pochammer's symbol)
and the falling factorial.
It is a special case of polynomial sequence of the binomial type studied in combinatorics theory.
In terms of the gamma function, an extension is defined for negative integers and even complex values.
Properties, mainly composition laws and binomial formulae, are given. They are used to evaluate
families of generalized Vandermonde determinants with $s$-shifted factorials as elements,
instead of power functions.
\par}

\bigskip
\bigskip
\noindent{\bf 1. Introduction}

\bigskip
\noindent This work has been motivated by studies of the probability density of the determinant (PDD)
of random matrices [1--3].
The method used, and sketched in section 5, is to compute the Mellin transform of the PDD.
In many cases it turned out to be a determinant with gamma functions as elements.
One aim of this work is to evaluate some of these
determinants and more generally determinants with shifted factorials (or Pochhammer's symbols) as elements.

We define in section 2 the $s$-shifted factorial $(z)_{s;n}$, equation (2.1), as a generalization for
non-negative values of $n$ of the power function $z^n$, the rising factorial $(z)_n$, equation (2.3), and
the falling factorial $[z]_n$, equation (2.4); both the names and the notations of these last two objects
are not well established, see [4--10]~
\footnote{${}^1$}{See [4] Pochhammer's symbol $(z)_n:=z(z+1)\cdots(z+n-1)$ 6.1.22,
[5] $(z)_n:=z(z+1)\cdots(z+n-1)$ p. xLiii,
[6] `{\it factorielle} $z$ {\it descendante d'ordre} $n$' $(z)_n:=z(z-1)\cdots(z-n+1)$ [4f],
`{\it factorielle} $z$ {\it montante d'ordre} $n$' or Pochhammer's symbol $n$ $<z>_n:=z(z+1)\cdots(z+n-1)$ [4g],
[7] {\it lower factorial} $(z)_n:=z(z-1)\cdots(z-n+1)$ (1.1),
{\it upper factorial} $z^{(n)}:=z(z+1)\cdots(z+n-1)$ (1.2),
[8] section 5 {\it falling factorial sequence} $(z)_n:=z(z-1)\cdots(z-n+1)$ 2.1,
{\it rising factorial sequence} $<z>_n:=z(z+1)\cdots(z+n-1)$ 3.1,
[9] {\it falling factorial of length} $n$ $[z]_n:=z(z-1)\cdots(z-n+1)$ 3.2,
{\it rising factorial of length} $n$ $[z]^n:=z(z+1)\cdots(z+n-1)$ 3.4 and III.2.A
and [10] $z^{\underline n}:=z(z-1)\cdots(z-n+1)$ $n${\it th falling power of} $z$ and
$z^{\overline n}:=z(z+1)\cdots(z+n-1)$ $n${\it th rising power of} $z$ 3.4.2.}
.
As a function of $z$, the $s$-shifted factorial is a special case of the polynomial sequences of the binomial type
studied mainly in the calculus of finite differences and combinatorics;
see in particular [6, 7] and for a wide bibliography [8, 9].
Expressed in terms of gamma functions, the $s$-shifted factorial can be extended to negative values
and even complex values of $n$.
The $s$-shifted factorial provides a compact formulation which emphasizes similarities and connections
which exist between the power function and the shifted factorials:
multiplication laws, Pascal triangle property, generating function and binomial formulae.

It is shown in section 3 that Vandermonde determinant with $(z_j)_{s;i}$ instead of $(z_j)^i$
as elements is still equal to the usual Vandermonde determinant. Other determinants with the inverse
of a $s$-shifted factorial, or the ratio of two $s$-shifted factorials, as elements are also evaluated,
both for positive and negative values of the index $i$.
Using the relations between the $s$-shifted factorial and the gamma function, or the binomial coefficient,
to each determinant evaluated in section 3, it corresponds a determinant in section 4 with elements expressed in
terms of gamma functions.
Finally, some applications of these determinants are given in section 5:
evaluation of the PDD of random matrices and also, a possible application to Stieltjes moment problems
arising in connection with the boson normal ordering problem.
As another example of application of the binomial formula, the finite sum of $s$-shifted factorials
of arithmetic progression to n terms is evaluated in appendix A.
Some basic properties of the product of differences and of the Vandermonde determinant are recalled,
respectively, in appendices B and C. Finally, appendix D illustrates another way to handle $s$-shifted
factorials.

\medskip
\bigskip
\noindent {\bf 2. Definition and some properties of the $s$-shifted factorial}

\bigskip
\noindent{\it 2.1. Definitions and relations between shifted factorials}

\smallskip
\noindent With $n$ a non-negative integer, $z$ and $s$ (the {\it shift}) some complex numbers, let us define the
$s$-{\it shifted factorial} by
$$(z)_{s;n}:=\cases{
1 &$n=0$ \cr
\noalign{\smallskip}
z(z+s)\cdots\bigl(z+(n-1)s\bigr) &$n=1,2,\ldots\,$. \cr
} \eqno(2.1)$$
For $s=0$, $1$ and $-1$, this definition coincides, respectively, with the power function,
the {\it rising factorial} (or {\it Pochhammer's symbol}, mainly in hypergeometric theory)
and the {\it falling factorial}, namely for $n$ nonzero,
$$\eqalignno{
(z)_{0;n} &=z^n &(2.2)\cr
(z)_{1;n} &=(z)_n:=z(z+1)\cdots(z+n-1) &(2.3)\cr
(z)_{-1;n} &=[z]_n:=z(z-1)\cdots(z-n+1) &(2.4)\cr
}$$
and when $n=0$ all these quantities take the value $1$.
Thereby, the $s$-shifted factorial allows compact expressions which emphasize the similarities
between the power function and the shifted factorials.

For any non-negative integer $n$, one has
$$\eqalignno{
(z)_{s;n} &=(-1)^n\,(-z)_{-s;n} &(2.5)\cr
&=\bigl(z+(n-1)s\bigr)_{-s;n}\,. &(2.6)\cr
}$$
For $s$ nonzero, the $s$-shifted factorials are related to the rising factorial, equation (2.3), by
$$(z)_{s;n}=s^n\,\Bigl({z\over s}\Bigr)_n\,. \eqno(2.7)$$

As a function of $z$, $(z)_{s;n}$ is a monic polynomial (i.e. the
coefficient of the highest power is one) in $z$ of degree $n$,
$$(z)_{s;n}=z^n+{n(n-1)\over2}\,s\,z^{n-1}+\cdots+(n-1)!\,s^{n-1}\,z \eqno(2.8)$$
with $0,-s,\ldots,-(n-1)\,s$ as zeros.
Consequences of these properties in terms of Vandermonde determinants are developed in section 3.
The sets of polynomials $\{(z)_{s;n}, n=0,1,\ldots\}$ are special cases
of the {\it polynomial sequences}
$\{{\rm p}_n(z), n=0,1,\ldots\}$, ${\rm p}_n(z)$ being exactly of degree $n$.
We are going to use these sequences in the way it is done in combinatorics [7--9]
\footnote{${}^2$}{See, e.g., [7] section 1, [8] section 3, [9] section III.2.}.
Any polynomial sequence is a basis of the vector space ${\cal P}$
over the complex field of complex polynomials in the variable $z$.
Then, to any two polynomial sequences $\{{\rm p}_n(z)\}$ and $\{{\rm q}_n(z)\}$
there exist uniquely determined {\it connecting coefficients} such that
${\rm q}_n(z)=\sum_{k=0}^n c_{n,k}\,{\rm p}_k(z)$.
These important coefficients have been widely studied, e.g.,
$$[z]_n=\sum_{k=0}^n s(n,k)\,z^k\qquad
z^n=\sum_{k=0}^n S(n,k)\,[z]_k\qquad
(z)_n=\sum_{k=0}^n L(n,k)\,[z]_k \eqno(2.9)$$
where $s(n,k)$, $S(n,k)$ and $L(n,k)={n-1\choose k-1}n!/k!$ are, respectively, the Stirling numbers
of the first and second kind [4, 6, 7, 9, 10]
\footnote{${}^3$}{See, e.g., [4] 24.1.3,4, [6] chapter V [5e,f], [7] (1.11--13), [9] 3.24,25 or [10] 2.5.2.}
and the signless Lah numbers [6, 7, 9, 10]
\footnote{${}^4$}{See, e.g., [6] chapter III p. 165 , [7] (1.14) and section 9, [9] 3.24,25, [10] 3.1.8.}
(other relations between $z^n$, $(z)_n$ and $[z]_n$ immediately follow from equations (2.5) and (2.6)).
We will see in subsection 2.6 that $\{(z)_{s;n}, n=0,1,\ldots\}$ has in addition the important property
to be a polynomial sequence of the binomial type.

\bigskip
\noindent{\it 2.2. Special values}

\smallskip
\noindent With $k$ some non-negative integer, one gets
$$\eqalignno{
(-k)_n=(-1)^n\,[k]_n &=\cases{
0 &$k=0,\ldots,n-1$ \cr
\noalign{\smallskip}
\displaystyle{(-1)^n{k!\over(k-n)!}} &$k=n,n+1,\ldots$ \cr} &(2.10)\cr
\noalign{\smallskip}
(k)_n=(-1)^n\,[-k]_n &=\cases{
0 &$k=0$ \cr
\noalign{\smallskip}
\displaystyle{{(k+n-1)!\over(k-1)!}} &$k=1,2,\ldots\,$. \cr} &(2.11)}$$
Then, for $s$ nonzero, values of $(ks)_{s;n}$ follow from equation (2.7), in particular $(s)_{s;n}=n!s^n$.

\bigskip
\noindent{\it 2.3. Relations with the gamma function and definition of the generalized $s$-shifted factorial}

\smallskip
\noindent One has [4]
\footnote{${}^5$}{See, e.g., [4] 6.1.5, 6.1.21 and 6.1.22.}
$$\eqalignno{
(z)_n &={\Gamma(z+n)\over\Gamma(z)}={(z+n-1)!\over(z-1)!}=(-1)^n\,n!\,{-z\choose n} &(2.12)\cr
[z]_n &={\Gamma(z+1)\over\Gamma(z-n+1)}={z!\over(z-n)!}=n!\,{z\choose n}\,. &(2.13)\cr
}$$
For $s$ nonzero, relations with $(z)_{s;n}$ follow from equation (2.7).

Actually, the relations above can be taken as the definition of $(z)_{s;n}$ in terms of the gamma function.
Thereby, one extends the $s$-shifted factorial to negative values, and even to complex values $t$ of $n$,
defining the {\it generalized $s$-shifted factorial} by
$$(z)_{s;t}:=s^t\,{\Gamma\bigl({z\over s}+t\bigr)\over\Gamma\bigl({z\over s}\bigr)} \eqno(2.14)$$
with a cut, say, along the negative real axis of the complex $s$ plane, with $-\pi<{\arg}\,s\le\pi$,
to ensure a single-valued dependence on $s$, and we choose the determination such that $s^t=1$ for $s=1$.
As $s$ goes to zero, say, along the real axis, using the Stirling formula [4]
\footnote{${}^6$}{See, e.g., [4] 6.1.37.},
$\Gamma(z)\sim e^{-z}z^{z-{1\over2}}(2\pi)^{1\over2}\bigl(1+O({1/z})\bigr)$
as $z\rightarrow\infty$ in $\vert{\rm arg}\,z\vert<\pi$,
one recovers the function $z^t$.
Furthermore, the definitions of the rising and falling factorials are extended by
$$(z)_t:=(z)_{1;t}\qquad[z]_t:=(z)_{-1;t}\,. \eqno(2.15)$$
Then, it immediately follows from the definition (2.14),
$$\eqalignno{
(z)_{s;0} &=1 &(2.16)\cr
(z)_{s;t} &={s^t\over(-s)^t}\,(-z)_{-s;t} &(2.17)\cr
}$$
by the recurrence formula $\Gamma(z+1)=z\,\Gamma(z)$,
$$(z)_{s;1}=z \eqno(2.18)$$
and by the reflection formula [4]
\footnote{${}^7$}{See, e.g., [4] 6.1.17.},
$\Gamma(z)\,\Gamma(1-z)=\pi/\sin(\pi z)$,
$$(z)_{s;t}={s^t\,\sin\bigl(\pi{z\over s}\bigr)\over(-s)^t\,\sin\bigl(\pi({z\over s}+t)\bigr)}\,
\bigl(z+(t-1)s\bigr)_{-s;t}\,. \eqno(2.19)$$
Thus  for any integer $q$,
$$\eqalignno{
(z)_{s;q} &=(-1)^q\,(-z)_{-s;q} &(2.20)\cr
&=\bigl(z+(q-1)s\bigr)_{-s;q} &(2.21)\cr
}$$
which generalize equations (2.5) and (2.6) for any integer, even negative.

\bigskip
\noindent{\it 2.4. Multiplication laws}

\smallskip
\noindent When the power function fulfils $z^t\,z^r=z^{t+r}$ with $r$ and $t$ some complex numbers,
it follows from the definition (2.14),
$$(z)_{s;t}\,(z+ts)_{s;r}=(z)_{s;t+r} \eqno(2.22)$$
and in particular, by equation (2.16), setting $r=-t$ yields
$$(z)_{s;t}={1\over(z+ts)_{s;-t}}\,. \eqno(2.23)$$
This relation generalizes $z^t=1/z^{-t}$ for $s=0$ and provides the relation between the
$s$-shifted factorials for any integer $q$ and $-q$, by equation (2.21),
$$(z)_{s;-q}={1\over(z-qs)_{s;q}}={1\over(z-qs)\bigl(z-(q-1)s\bigr)\cdots(z-s)}
={1\over(z-s)_{-s;q}}\,. \eqno(2.24)$$
In terms of binomial coefficients, the multiplication law (2.22) reads
$${z\choose n}{z-n\choose p}={n+p\choose n}{z\choose n+p}
\qquad{\rm or}\qquad
{z\choose n}\,[n]_p=[z]_p\,{z-p\choose n-p}\,. \eqno(2.25)$$

\smallskip
For a proper choice of determination the power function fulfils $w^tz^t=(wz)^t$.
For the $s$-shifted factorial, one has
$$(wz)_{s;t}={s^t\over({s\over w})^t}\,(z)_{{s\over w};t} \eqno(2.26)$$
and thus for any integer $q$,
$$(wz)_{s;q}=w^q\,(z)_{{s\over w};q}\,. \eqno(2.27)$$
For $w=-1$, relation (2.26) corresponds to equation (2.17).
For $w=k$ and $q=n$ some non-negative integers, iterating the multiplication law (2.22) and from equation (2.27),
$$(kz)_{s;kn}=k^{kn}\prod_{\ell=0}^{k-1}\prod_{j=0}^{n-1}\bigl(z+(n\ell+j)\,{s\over k}\bigr)
=k^{kn}\prod_{\ell=0}^{k-1}\bigl(z+{\ell\over k}\,s\bigr)_{s;n} \eqno(2.28)$$
where the last equality corresponds to a rearrangement of the factors, both $n\ell+j$ and $\ell+jk$
taking once all the $kn$ values $0,1,\ldots,kn-1$.
The equation above can also be obtained from the definition (2.14) and the Gauss multiplication formula [4]
\footnote{${}^8$}{See, e.g., [4] 6.1.20.}
,
$\Gamma(kz)=(2\pi)^{{1\over2}(1-k)}k^{kz-{1\over2}}\prod_{\ell=0}^{k-1}
\Gamma\bigl(z+{\ell/k}\bigr)$.
Note that, based on the reflection formula and the Gauss' multiplication formula,
$2\sin(\pi kz)$ follows the known multiplication law similar to equation (2.28) [5]
\footnote{${}^9$}{See, e.g., [5] 1.392 (1.).}
,
$$2\,\sin(\pi kz)=\prod_{\ell=0}^{k-1}2\,\sin\Bigl(\pi\bigl(z+{\ell\over k}\bigr)\Bigr)\,. \eqno(2.29)$$

\smallskip
For a proper choice of determination the power function fulfils $(z^t)^r=z^{tr}$.
No equivalent general relation exists for the $s$-shifted factorial.
Although $(z^{-1})_p$ has no simple relation with $((z)_p)^{-1}$,
let us point out the following expression for any integers $n\ge p\ge0$,
by equations (2.24), (2.22) and (2.6), with $z\ne0,-s,\ldots,-(n-1)s$,
$${1\over(z)_{s;p}}=(z+ps)_{s;-p}
={(z+ps)_{s;n-p}\over(z)_{s;n}}={\bigl(z+(n-1)s\bigr)_{-s;n-p}\over(z)_{s;n}} \eqno(2.30)$$
recovering for $s=0$ the relation $(z^p)^{-1}=z^{-p}=z^{n-p}\,(z^n)^{-1}$.

\bigskip
\noindent{\it 2.5. Generalized Pascal triangle property and $s$-difference operator}

\smallskip
\noindent The multiplication law (2.22) and equation (2.18) yield
$$(z)_{s;t}-(z-s)_{s;t}=ts\,(z)_{s;t-1} \eqno(2.31)$$
which generalizes the Pascal triangle property for binomial coefficients, by equation (2.12),
$${z+1\choose n}={z\choose n}+{z\choose n-1}\,. \eqno(2.32)$$

Let us define the $s$-{\it difference operator} $\Delta_s$ on functions ${\rm f}$ of $z$ by
$$\Delta_s {\rm f}(z):={\rm f}(z+s)-{\rm f}(z) \eqno(2.33)$$
(this operator must not be confused with the product of differences
$\Delta_n({\bf z})$ introduced latter in section 3 and defined by equation (B.1)).
It follows immediately from equation (2.31),
$$\Delta_s(z)_{s;t}=ts\,(z+s)_{s;t-1}\qquad\Delta_{-s}(z)_{s;t}=-ts\,(z)_{s;t-1} \eqno(2.34)$$
and iterating these formulae, e.g., the first one
$$\Delta_s^p\,(z)_{s;t}=[t]_p\,s^p\,(z+ps)_{s;t-p} \eqno(2.35)$$
recovering for $s=0$ the expression of ${d^p\over dz^p}z^n$.

\bigskip
\noindent{\it 2.6. Generating function and binomial formulae}

\smallskip
\noindent With $x$ some complex variable, let ${\rm G}_{s;z}(x)$ be the generating function of the
$s$-shifted factorials $(z)_{s;n}$,
$${\rm G}_{s;z}(x):=\sum_{n=0}^\infty(z)_{s;n}\,{x^n\over n!}\,,\quad\vert s x\vert<1 \eqno(2.36)$$
and, using equation (2.7),
$${\rm G}_{s;z}(x)={\rm G}_{1;{z\over s}}(s x)\,. \eqno(2.37)$$
Now, the generating function of the rising factorials can be obtained directly from
the binomial series with equation (2.12),
$$(1-x)^{-z}=\sum_{n=0}^\infty(-1)^n\,{-z\choose n}\,x^n
=\sum_{n=0}^\infty(z)_n\,{x^n\over n!}={\rm G}_{1;z}(x)\,,\quad\vert x\vert<1\,. \eqno(2.38)$$
Therefore,
$${\rm G}_{s;z}(x)=(1-s x)^{-z\over s} \eqno(2.39)$$
recovering for $s=0$ the expression ${\rm G}_{0,z}(x):=\sum_{n=0}^\infty z^n\,{x^n\over n!}=e^{xz}$.

\smallskip
Since the generating function ${\rm G}_{s;z}(x)$ above reads as
an exponential function ${\rm F}(x)^z$ of $z$, it satisfies the multiplication law [11]
$${\rm G}_{s;z}(x)\,{\rm G}_{s;w}(x)={\rm G}_{s;z+w}(x)\,. \eqno(2.40)$$
Expanding both sides of this last equation as a power series in $x$ yields,
$$(z+w)_{s;n}=\sum_{k=0}^n{n\choose k}\,(z)_{s;k}\,(w)_{s;n-k} \eqno(2.41)$$
namely, the $s$-shifted factorial satisfies the binomial formula.
The polynomial sequence $\{(z)_{s;n},n=0,1,\ldots\}$ which satisfies $(z)_{s;0}=1$ and the binomial formula above
is said to be of {\it binomial type} [6--10]
\footnote{${}^{10}$}{See, e.g., [6] [6a] and [13c,d], [7] (1.6), [8] section 5, [9] section III.2, [10] 3.4.2 (3.).}.
This property is shared by many other {\it binomial sequences}
$\{{\rm p}_n(z),n=0,1,\ldots\}$
which have been studied mainly in combinatorics using generating function methods and above all efficient
operator methods.

The binomial sequences can be characterized by a generating function which depends exponentially on $z$ [9]
\footnote{${}^{11}$}{See, e.g., [9] 3.59.}
$${\rm G}_z(x)=e^{{\rm g}(x)z}=e^{(x+g_2x^2+\cdots)\,z}
=\sum_{n=0}^\infty {\rm p}_n^{\{{\rm g}\}}(z)\,{x^n\over n!} \eqno(2.42)$$
then ${\rm p}_n^{\{{\rm g}\}}(z)$ is a monic polynomial of degree $n$ in $z$,
the coefficients of which are known as {\it Bell polynomials} [6]
\footnote{${}^{12}$}{See, e.g., [6] section III.3.}
(indeed, expanding the exponential series above, the term in $z^n$ reads $z^n\,x^n\bigl(1+O(x)\bigr)/n!$).
In the case we consider, ${\rm g}(x):=-s^{-1}\ln(1-sx)$ and ${\rm p}_n^{\{{\rm g}\}}(z)=(z)_{s;n}$;
for $s=0$, ${\rm g}(x)=x$ and ${\rm p}_n^{\{{\rm g}\}}(z)=z^n$.
The binomial sequences can as well be characterized by the fact [7--9]
\footnote{${}^{13}$}{See, e.g., [7] section 3 theorem 1, [8] section 7, [9] 3.45.}
that the {\it basis operator} of the sequence, i.e. the linear operator D
of the vector space ${\cal P}$ (already considered in subsection 2.1) into itself
defined by $D{\rm p}_0(z):=0$ and $D{\rm p}_n(z):=n{\rm p}_{n-1}(z)$ for $n\ge1$,
is a {\it delta operator}, i.e. it is shift invariant, $DE_a=E_aD$ for all complex number $a$,
where $E_a$ is the {\it translation operator} defined by $E_a{\rm f}(z):={\rm f}(a+z)$
and moreover $Dz=c\ne0$.
In our case, from equation (2.34), $D_s:=-\Delta_{-s}/s=({\bf I}-E_{-s})/s$,
where ${\bf I}$ is the identity operator, is clearly shift invariant and $D_sz=1$.
Indeed, the binomial formula (2.41) for the $s$-shifted factorial can also be demonstrated by recurrence
from the Pascal triangle properties (2.31) and (2.32).
It is true for $n=0$ and $1$. Let us assume it to be true for $n$, then,
$$\eqalignno{
(z+w)_{s;n+1} &=\sum_{k=0}^n{n\choose k}\,(z)_{s;k}\,(w)_{s;n-k}\,\bigl(z+ks+w+(n-k)s\bigr) \cr
&=\sum_{k=0}^{n+1}{n\choose k-1}\,(z)_{s;k}\,(w)_{s;n+1-k}+\sum_{k=0}^{n+1}{n\choose k}\,
(z)_{s;k}\,(w)_{s;n+1-k} \cr
&=\sum_{k=0}^{n+1}{n+1\choose k}\,(z)_{s;k}\,(w)_{s;n+1-k}\,. &(2.43)\cr
}$$

\smallskip
By equation (2.12), in terms of binomial coefficients, the binomial formula (2.41) reads
$${z+w\choose n}=\sum_{k=0}^n{z\choose k}{w\choose n-k}\,. \eqno(2.44)$$
As for the power function, the binomial formula for the $s$-shifted factorial can be directly extended to
$p>2$ variables using multinomial coefficients,
$$\Bigl(\sum_{j=1}^p z_j\Bigr)_{s;n}=\sum_{n_1,\ldots,n_p=0\atop n_1+\cdots+n_p=n}^n
{n!\over n_1!\cdots n_p!}\,(z_1)_{s;n_1}\cdots(z_p)_{s;n_p}\,. \eqno(2.45)$$
From equation (2.5) the following corollary is immediately obtained:
$$(z-w)_{s;n}=\sum_{k=0}^n(-1)^{n-k}{n\choose k}(z)_{s;k}\,(w)_{-s;n-k}\,. \eqno(2.46)$$

Although, as already noted, $(z^{-1})_i\ne((z)_i)^{-1}$, the binomial formula can be extended to
the inverse of $s$-shifted factorials. Indeed, by equations (2.30), (2.41) and (2.6),
$$\eqalignno{
\sum_{k=0}^n{n\choose k}\,{1\over(z)_{s;k}}\,{1\over(w)_{s;n-k}}
&=\sum_{k=0}^n{n\choose k}\,{\bigl(z+(n-1)s\bigr)_{-s;n-k}\over(z)_{s;n}}\,
{\bigl(w+(n-1)s\bigr)_{-s;k}\over(w)_{s;n}} \cr
&={\bigl(z+w+2(n-1)s\bigr)_{-s;n}\over(z)_{s;n}\,(w)_{s;n}}
={\bigl(z+w+(n-1)s\bigr)_{s;n}\over(z)_{s;n}\,(w)_{s;n}} &(2.47)\cr
}$$
corresponding for $s=0$ to
$$\sum_{k=0}^n{n\choose k}\,{1\over z^k}\,{1\over w^{n-k}}
={(z+w)^n\over z^n\,w^n}=\biggl({1\over z}+{1\over w}\biggr)^n \eqno(2.48)$$
where, once again, the last equality above does not hold for $s$ nonzero.
Similarly, one also gets from equations (2.30) and (2.46)
$$\eqalignno{
\sum_{k=0}^n(-1)^{n-k}{n\choose k}{(z)_{s;k}\over(w)_{s;k}}
&=\sum_{k=0}^n(-1)^{n-k}{n\choose k}\,(z)_{s;k}\,{\bigl(w+(n-1)s\bigr)_{-s;n-k}\over(w)_{s;n}} \cr
&={\bigl(z-w-(n-1)s\bigr)_{s;n}\over(w)_{s;n}} &(2.49)\cr
}$$
corresponding for $s=0$ to
$$\sum_{k=0}^n(-1)^{n-k}{n\choose k}{z^k\over w^k}
={(z-w)^n\over w^n}=\biggl({z\over w}-1\biggr)^n\,\,. \eqno(2.50)$$

Another kind of useful relations is as follows. From equation (2.25), the multiplication law (2.22)
and the binomial formula (2.41),
$$\eqalignno{
\sum_{k=0}^n{n\choose k}\,[k]_p\,(z)_{s;k}\,(w)_{s;n-k}
&=[n]_p\,\sum_{k=p}^n{n-p\choose k-p}\,(z)_{s;k}\,(w)_{s;n-k} \cr
&=[n]_p\,(z)_{s;p}\,\sum_{k=p}^n{n-p\choose k-p}\,(z+ps)_{s;k-p}\,(w)_{s;n-k} \cr
&=[n]_p\,(z)_{s;p}\,(z+w+ps)_{s;n-p}\,. &(2.51)\cr
}$$

Several of these binomial formulae are used in the next section to evaluate some determinants
with $s$-shifted factorials as elements.
As another example of application, the finite sum of
$s$-shifted factorials of arithmetic progression to $n$ terms is evaluated in appendix A.

\medskip
\bigskip
\noindent{\bf 3. Generalized Vandermonde determinant with $s$-shifted factorials as elements}

\bigskip
\noindent In what follows, $n$ is a positive integer and ${\bf z}$ is,
either a set of complex numbers, or a complex function,
$${\bf z}:=\{z_j,j=0,\ldots,n-1\}\qquad{\rm or}\qquad j\mapsto{\bf z}(j):=z_j\quad j=0,\ldots,n-1\,. \eqno(3.1)$$
Some basic properties of the product of differences $\Delta_n({\bf z}):=\prod_{0\le i<j\le n-1}(z_j-z_i)$,
equation (B.1), and of the Vandermonde determinant $\det\,\bigl[(z_j)^i\bigr]_{i,j=0,\ldots,n-1}$,
are recalled, respectively, in appendices B and C.

\bigskip
\noindent {\it 3.1. Expressions for $s$-shifted factorial with a non-negative index}

\smallskip
\noindent {\bf Lemma 1.}
{\sl
With $n$ a positive integer and $s$ some complex number,
the {\it generalized Vandermonde determinant of $s$-shifted factorials}, still is the product of differences,
$$\det\,\bigl[(z_j)_{s;i}\bigr]_{i,j=0,\ldots,n-1}=\Delta_n({\bf z}) \eqno(3.2)$$
thus, it does not depends on $s$. More generally,
$$\det\bigl[\Pi_i(z_j)\bigr]_{i,j=0,\ldots,n-1}=\lambda\,\Delta_n({\bf z}) \eqno(3.3)$$
where $\Pi_i(z)$ are $n$ linearly independent {\it polynomials in} $(z)_{s;.}$
{\it each of degree less than} $n$ and defined as follows:
$$\Pi_i(z):=\sum_{k=0}^{n-1} c_{i,k}\,(z)_{s;k}\quad i=0,\ldots,n-1\qquad
\lambda:=\det\bigl[c_{i,k}\bigr]_{i,k=0,\ldots,n-1}\ne0\,. \eqno(3.4)$$
In particular, with $b_i$ some complex numbers, one has
$$\det\,\bigl[(b_i+z_j)_{s;i}\bigr]_{i,j=0,\ldots,n-1}=\Delta_n({\bf z})\,. \eqno(3.5)$$
Finally, with $t$ some complex number,
$$\det\,\bigl[(z_j)_{s;t+i}\bigr]_{i,j=0,\ldots,n-1}
=\Bigl\{\prod_{j=0}^{n-1}(z_j)_{s;t}\Bigr\}\,\Delta_n({\bf z})\,. \eqno(3.6)$$
}

\medskip
Two proofs are given. Based on the properties of the $s$-shifted factorial,
proof 1 expresses the determinants considered in terms of Vandermonde determinants.
Illustrating again the similarities between $(z)_{s;i}$ and $z^i$, proof 2 follows the same steps as
a usual way of computing the Vandermonde determinant, equation (C.1).

\medskip
\noindent {\bf Proof 1.}
The $s$-shifted factorial $(z)_{s;i}$ is a monic polynomial of degree $i$ in $z$, see equation (2.8).
Hence, formula (3.2) follows from equation (C.3).
Note that equation (3.2) still holds for the monic polynomials obtained from any generating function
defined by equation (2.42).
Formulae (3.3) and (3.5) can be directly obtained either from equation (C.3) in terms
of usual polynomials (e.g., $(b_i+z)_{s;i}$ is also a monic polynomial of degree $i$ in $z$)
or starting from formula (3.2), by the same arguments as for equation
(C.3), in terms of polynomials in $s$-shifted factorials (e.g., by the binomial formula (2.41),
$(b_i+z)_{s;i}=\sum_{k=0}^i{i\choose k}(b_i)_{s;k}(z)_{s;n-k}$,
i.e. a monic polynomial of degree $i$ in $(z)_{s;.}$).
Finally, equation (3.6) follows from the multiplication law (2.22) and formula (3.5).

\medskip
\noindent {\bf Proof 2.}
Let $M_{i,j}:=(z_j)_{s;i}$.
The determinant $\det[M_{i,j}]_{i,j=0,\ldots,n-1}$ is not changed if one replaces the row ${\cal R}_i$
by the linear combination ${\cal R}_i-(M_{i,0}/M_{i-1,0})\,{\cal R}_{i-1}$,
successively for $i=n-1,n-2,\ldots,1$.
Then, by the multiplication law (2.22), for $i=1,\ldots,n-1$ and $j=0,\ldots,n-1$,
$$M_{i,j}\quad\rightarrow\quad(z_j)_{s;i}-\bigl(z_0+(i-1)s\bigr)\,(z_j)_{s;i-1}
=(z_j-z_0)\,(z_j)_{s;i-1}\,. \eqno(3.7)$$
This operation replaces the column ${\cal C}_0$ by zeros except for the row ${\cal R}_0$ left unchanged.
Expanding the determinant with respect to ${\cal C}_0$ and taking out the
factors depending only on $j$ yield the recurrence formula on $n$,
$$\eqalignno{
D_{s;n}(z_0,\ldots,z_{n-1}) &:=\det\bigl[M_{i,j}\bigr]_{i,j=0,\ldots,n-1} \cr
&\phantom{:}=\Bigl\{\prod_{j=1}^{n-1}(z_j-z_0)\Bigr\}\,D_{s;n-1}(z_1,\ldots,z_{n-1})\,. &(3.8)\cr
}$$
An iteration of this last equation, down to $D_{s;1}(z_{n-1})=1$, completes the proof.

\smallskip
The recurrence procedure above makes step by step the matrix $(M_{i,j})_{i,j=0,\ldots,n-1}$ triangular.
Let us denote by a superscript the rank of the step in this procedure.
At the first step the row $i=0$ is unchanged while for $i=1,\ldots,n-1$,
$${\cal R}_i^{(1)}={\cal R}_i-{M_{i,0}\over M_{i-1,0}}\,{\cal R}_{i-1}\,. \eqno(3.9)$$
At the second step the rows $i=0,1$ are unchanged, while for $i=2,\ldots,n-1$,
$$\eqalignno{
{\cal R}_i^{(2)} &={\cal R}_i^{(1)}-{M_{i,1}^{(1)}\over M_{i-1,1}^{(1)}}\,{\cal R}_{i-1}^{(1)} \cr
&={\cal R}_i-\biggl({M_{i,0}\over M_{i-1,0}}+{M_{i,1}^{(1)}\over M_{i-1,1}^{(1)}}\biggr)\,{\cal R}_{i-1}
+{M_{i,1}^{(1)}\over M_{i-1,1}^{(1)}}\,{M_{i-1,0}\over M_{i-2,0}}\,{\cal R}_{i-2}\,. &(3.10)
}$$
Generically, the final expression of the row $i$ is given by ${\cal R}_i^{(i)}$.
It happens that in the special case $z_j:=b+js$, with $b$ some complex number and $s$ nonzero,
these expressions read
$$\eqalignno{
{\cal R}_i^{(1)} &={\cal R}_i-\bigl(b+(i-1)s\bigr)\,{\cal R}_{i-1} &(3.11)\cr
{\cal R}_i^{(2)} &={\cal R}_i-2\bigl(b+(i-1)s\bigr)\,{\cal R}_{i-1}
+\bigl(b+(i-1)s\bigr)\bigl(b+(i-2)s\bigr)\,{\cal R}_{i-2} &(3.12)\cr
&\vdots \cr
{\cal R}_i^{(i)} &=\sum_{k=0}^i(-1)^{i-k}{i\choose k}\,\bigl(b+(i-1)s\bigr)_{-s;i-k}\,{\cal R}_k &(3.13)\cr
}$$
where the last formula can be checked as follows.
With $M_{i,j}:=(b+js)_{s;i}$, by the binomial formula (2.46), equations (2.7) and (2.6),
$$\eqalignno{
M_{i,j}^{(i)} &=\sum_{k=0}^i(-1)^{i-k}{i\choose k}\,\bigl(b+(i-1)s\bigr)_{-s;i-k}\,(b+js)_{s;k} \cr
&=s^i\,(j-i+1)_i=s^i\,[j]_i &(3.14)\cr
}$$
which vanishes for $i>j$, see equation (2.10).
Another proof of this identity is given in appendix D.1.
Thus, as expected, the resulting matrix is triangular
and its determinant is the product of its diagonal elements $s^j\,[j]_j=s^j\,j!$. Then, by equation (B.5),
$$\det\,\bigl[(b+js)_{s;i}\bigr]_{i,j=0,\ldots,n-1}=s^{n(n-1)/2}\prod_{j=0}^{n-1}j!
=\Delta_n(j\mapsto b+js) \eqno(3.15)$$
completing the proof of equation (3.2) in the special case $z_j:=b+js$.

\bigskip
\noindent {\bf Lemma 2.}
{\sl
With $n$ a positive integer, $s$ some complex number and $z_j\ne0,-s,\ldots,-(n-2)s$,
$$\det\biggl[{1\over(z_j)_{s;i}}\biggr]_{i,j=0,\ldots,n-1}
={(-1)^{n(n-1)/2}\over\prod_{j=0}^{n-1}(z_j)_{s;n-1}}\ \Delta_n({\bf z})\,. \eqno(3.16)$$
}
This formula generalizes equation (C.5).

\medskip
\noindent {\bf Proof 1.}
With $n-1\ge i\ge0$ and $z_j\ne0,-s,\ldots,-(n-2)s$,
by equations (2.30) one gets
$$\det\biggl[{1\over(z_j)_{s;i}}\biggr]_{i,j=0,\ldots,n-1}
={\det\bigl[\bigl(z_j+(n-2)s\bigr)_{-s;n-1-i}\bigr]_{i,j=0,\ldots,n-1}\over\prod_{j=0}^{n-1}(z_j)_{s;n-1}}\,.
\eqno(3.17)$$
Then, changing $i$ into $n-1-i$ (i.e. rearranging the rows) on the right-hand side determinant above,
lemma 2 follows from equation (3.5).
Note that when $s=0$, equation (C.5) for the power function can also be derived
as above from $1/z^i=z^{n-1-i}/z^{n-1}$.

\medskip
\noindent {\bf Proof 2.}
This proof of lemma 2 follows the same steps as proof 2 of lemma 1.
With the linear combination of rows ${\cal R}_i-(M_{i,0}/M_{i-1,0})\,{\cal R}_{i-1}$,
$$M_{i,j}:={1\over(z_j)_{s;i}}\quad\rightarrow\quad{-(z_j-z_0)\over\bigl(z_0+(i-1)s\bigr)\,z_j}\,
{1\over(z_j+s)_{s;i-1}}\,,
\quad i=1,\ldots,n-1\,. \eqno(3.18)$$
Then, with $z_j\ne0,-s,\ldots,-(n-2)s$, the recurrence formula on $n$ reads
$$\eqalignno{
D_{s;n}(z_0,\ldots,z_{n-1}) &:=\det\bigl[M_{i,j}\bigr]_{i,j=0,\ldots,n-1} \cr
&\phantom{:}={(-1)^{n-1}\,\prod_{j=1}^{n-1}(z_j-z_0)\over(z_0)_{s;n-1}\,\prod_{j=1}^{n-1}z_j}\,
D_{s;n-1}(z_1+s,\ldots,z_{n-1}+s)\,. &(3.19)\cr
}$$
Iteration of this last equation, down to $D_{s;1}\bigl(z_{n-1}+(n-1)s\bigr)=1$, ends the proof.

\smallskip
As in proof 2 of lemma 1, in the special case $z_j:=b+js$ with $b$ some complex number and $s$ nonzero,
the determinant can be made triangular in one step, replacing ${\cal R}_i$ by the linear combination
$${\cal R}_i^{(i)}=\sum_{k=0}^i{i\choose k}{1\over\bigl(-b-2(i-1)s\bigr)_{s;i-k}}\,{\cal R}_k\,. \eqno(3.20)$$
Indeed, with $M_{i,j}:=1/(b+js)_{s;i}$ and $b\ne0,-s,\ldots,-(2n-3)s$,
it follows from the binomial formula (2.47) and equations (2.5)--(2.7)
$$M_{i,j}^{(i)}=\sum_{k=0}^i{i\choose k}{1\over\bigl(-b-2(i-1)s\bigr)_{s;i-k}}\,{1\over(b+js)_{s;k}}
={(-s)^i\,[j]_i\over\bigl(b+(i-1)s\bigr)_{s;i}\,(b+js)_{s;i}} \eqno(3.21)$$
which vanishes for $i>j$.
Another proof of this identity is given in appendix D.2.
Then, the determinant is the product of its diagonal elements,
$$\det\biggl[{1\over(b+js)_{s;i}}\biggr]_{i,j=0,\ldots,n-1}
=(-s)^{n(n-1)/2}\prod_{j=0}^{n-1}{j!\over\bigl(b+(j-1)s\bigr)_{s;j}\,(b+js)_{s;j}} \eqno(3.22)$$
and, using the multiplication law (2.22), it can be shown by recurrence that for all $s$
$$\prod_{j=0}^{n-1}\bigl(b+(j-1)s\bigr)_{s;j}\,(b+js)_{s;j}=\prod_{j=0}^{n-1}(b+js)_{s;n-1} \eqno(3.23)$$
corresponding to a rearrangement of the factors.
Finally, by equation (B.5),
$$\det\biggl[{1\over(b+js)_{s;i}}\biggr]_{i,j=0,\ldots,n-1}
={(-1)^{n(n-1)/2}\over\prod_{j=0}^{n-1}(b+js)_{s;n-1}}\,\Delta_n(j\mapsto b+js) \eqno(3.24)$$
ending the proof of equation (3.16) in the special case $z_j:=b+js$.

\bigskip
\noindent {\bf Lemma 3.}
{\sl
With $n$ a positive integer, $a$ and $b$ some complex numbers and $az_j+b\ne0,-s,\ldots,-(n-2)s$,
$$\det\biggl[{(z_j)_{s;i}\over(az_j+b)_{s;i}}\biggr]_{i,j=0,\ldots,n-1}
=\biggl\{\prod_{j=0}^{n-1}{\bigl(b+(n-1-j)(1-a)s\bigr)_{s;j}\over(az_j+b)_{s;n-1}}\biggr\}\,\Delta_n({\bf z})
\,. \eqno(3.25)$$
}
This formula generalizes equation (C.6).

\medskip
\noindent {\bf Proof 1.}
From equation (2.30), with $n-1\ge i\ge0$ and $az+b\ne0,-s,\ldots,-(n-1)s$,
$${(z)_{s;i}\over(az+b)_{s;i}}={1\over(az+b)_{s;n-1}}\,(z)_{s;i}\,(az+b+is)_{s;n-1-i}\,. \eqno(3.26)$$
When $a=1$, from the binomial formula (2.41) and the multiplication law (2.22),
$$\eqalignno{
(z)_{s;i}\,(z+b+is)_{s;n-1-i}
&=\sum_{k=0}^{n-1-i}{n-1-i\choose k}(b)_{s;k}\,(z)_{s;i}\,(z+is)_{s;n-1-i-k} \cr
&=\sum_{k=0}^{n-1-i}{n-1-i\choose k}(b)_{s;k}\,(z)_{s;n-1-k}=\Pi_i(z) &(3.27)\cr
}$$
where $\Pi_i(z)$, a polynomial in $(z)_{s;.}$ of degree $n-1$, is defined as in equation (3.4) with
$$c_{i,k}:=\cases{
0 &$k=0,\ldots,i-1$\cr
\noalign{\smallskip}
\displaystyle{{n-1-i\choose n-1-k}\,(b)_{s;n-1-k}} &$k=i,\ldots,n-1\,$.\cr
} \eqno(3.28)$$
Thus, the matrix $[c_{i,k}]_{i,k=0,\ldots,n-1}$ is triangular and its determinant is the product
of its diagonal elements,
$$\det[c_{i,k}]_{i,k=0,\ldots,n-1}=\prod_{j=0}^{n-1}(b)_{s;j}\,. \eqno(3.29)$$
Then, when $a=1$, lemma 3 follows from equation (3.3), with $z_j+b\ne0,-s,\ldots,-(n-2)s$,
$$\eqalignno{
\det\biggl[{(z_j)_{s;i}\over(z_j+b)_{s;i}}\biggr]_{i,j=0,\ldots,n-1}
& =\biggl\{\,\prod_{j=0}^{n-1}{1\over(z_j+b)_{s;n-1}}\biggr\}\,\det[\Pi_i(z_j)]_{i,j=0,\ldots,n-1} \cr
&=\biggl\{\,\prod_{j=0}^{n-1}{(b)_{s;j}\over(z_j+b)_{s;n-1}}\biggr\}\,\Delta_n({\bf z})\,. &(3.30)\cr
}$$
Note that for $s=0$, the equation (C.6) for the power function can also be derived as above.

\noindent When $a\ne1$, $(z)_{s;i}\,(az+b+is)_{s;n-1-i}$ is still a polynomial $\Pi_i(z)$ of degree
$n-1$ in $(z)_{s;.}$ (or equivalently $z$). But now the evaluation of the connecting coefficients $c_{i,k}$
(or even to compute $\det[c_{i,k}]_{i,k=0,\ldots,n-1}$ we only need)
is no longer easy since there is no simple combination law between the $s$-shifted factorials of $z$ and $az$.
proof 2 provides a simple proof of lemma 3 for all values of $a$.

\medskip
\noindent {\bf Proof 2.}
This proof of equation (3.25) follows the same steps as proof 2 of lemma 1.
With the linear combinations of rows ${\cal R}_i-(M_{i,0}/M_{i-1,0})\,{\cal R}_{i-1}$,
$$M_{i,j}:={(z_j)_{s;i}\over(az_j+b)_{s;i}}\quad\rightarrow\quad
{(z_j-z_0)\bigl(b+(i-1)(1-a)s\bigr)\over\bigl(az_0+b+(i-1)s\bigr)\,(az_j+b)}\
{(z_j)_{s;i-1}\over(az_j+b+s)_{s;i-1}}\,,\quad i=1,\ldots,n-1\,. \eqno(3.31)$$
Then, with $az_j+b\ne0,-s,\ldots,-(n-2)s$, the recurrence formula on $n$ reads
$$\eqalignno{
D_{s;n}(z_0,\ldots,z_{n-1};a,b) &:=\det\bigl[M_{i,j}\bigr]_{i,j=0,\ldots,n-1} \cr
&\phantom{:}={(b)_{(1-a)s;n-1}\over(az_0+b)_{s;n-1}}\biggl\{\,\prod_{j=1}^{n-1}{z_j-z_0\over az_j+b}\biggr\}\,
D_{s;n-1}(z_1,\ldots,z_{n-1};a,b+s)\,. &(3.32)
}$$
Iteration of this equation, down to $D_{s;1}\bigl(z_{n-1};a,b+(n-1)s\bigr)=1$, ends the proof.

Note that in lemma 3, corresponding to a rearrangement of the factors,
$$\prod_{j=0}^{n-1}\bigl(b+(n-1-j)(1-a)s\bigr)_{s;j}
=\prod_{j=0}^{n-1}\bigl(b+(n-1-j)s\bigr)_{(1-a)s;j}\,. \eqno(3.33)$$

\smallskip
As in proof 2 of lemma 1, in the special case $z_j:=c+js$ with $c$ some complex number, $a=1$ and $s$ nonzero,
the determinant can be made triangular in one step, replacing ${\cal R}_i$ by the linear combination
$${\cal R}_i^{(i)}=\sum_{k=0}^i(-1)^{i-k}{i\choose k}
{\bigl(c+(i-1)s\bigr)_{-s;i-k}\over\bigl(d+2(i-1)s\bigr)_{-s;i-k}}\,{\cal R}_k \eqno(3.34)$$
where $d:=b+c$. Indeed, with $M_{i,j}:=(c+js)_{s;i}/(d+js)_{s;i}$ and $d\ne0,-s,\ldots,-(2n-3)s$,
after some elementary algebra based on the relations (2.5)--(2.7), (2.12) and (2.13), one gets
$$\eqalignno{
M_{i,j}^{(i)} &=\sum_{k=0}^i(-1)^{i-k}{i\choose k}
{\bigl(c+(i-1)s\bigr)_{-s;i-k}\over\bigl(d+2(i-1)s\bigr)_{-s;i-k}}\,{(c+js)_{s;k}\over(d+js)_{s;k}} \cr
&=(-1)^i\,{(c)_{s;i}\over\bigl(d+(i-1)s\bigr)_{s;i}}\,
{}_3{\rm F}_2(cs^{-1}+j,ds^{-1}+i-1,-i;cs^{-1},ds^{-1}+j;1) &(3.35)\cr
}$$
where the ${}_3{\rm F}_2$ is a terminating Saalsch\"utzian generalized hypergeometric series [12]
\footnote{${}^{14}$}{See, e.g., [12] equations 2.1(30) and 4.4(3).}
,
$${}_3{\rm F}_2(\alpha,\beta,-i;\gamma,1+\alpha+\beta-\gamma-i;1)
={(\gamma-\alpha)_i\,(\gamma-\beta)_i\over(\gamma)_i\,(\gamma-\alpha-\beta)_i}\quad i=0,1,\ldots\,. \eqno(3.36)$$
Hence,
$$M_{i,j}^{(i)}={s^i\,[j]_i\,(d-c)_{s;i}\over(d+js)_{s;i}\,\bigl(d+(i-1)s\bigr)_{s;i}} \eqno(3.37)$$
which vanishes for $i>j$.
Another proof of this identity is given in appendix D.3.
Then, the determinant is the product of its diagonal elements and with equation (3.23), one finds
$$\eqalignno{
\det\biggl[{(c+js)_{s;i}\over(d+js)_{s;i}}\biggr]_{i,j=0,\ldots,n-1}
&=\prod_{j=0}^{n-1}{s^j\,j!\,(d-c)_{s;j}\over(d+js)_{s;n-1}} \cr
&=\prod_{j=0}^{n-1}{(d-c)_{s;j}\over(d+js)_{s;n-1}}\,\Delta_n(j\mapsto c+js) &(3.38)\cr
}$$
ending the proof of equation (3.25) in the special case $z_j:=c+js$ and $a=1$.

\bigskip
\noindent {\it 3.2. Consequences for $s$-shifted factorial with a negative index}

\smallskip
\noindent Using equation (2.24), $(z)_{s;-i}=1/(z-s)_{-s;i}$, and (B.2),
the following corollaries are direct consequences of the previous lemmas.

\medskip
\noindent {\bf Corollary 1.}
{\sl
With $n$ a positive integer, $s$ some complex number and $z_j\ne s,2s,\ldots,(n-1)s$,
$$\det\,\bigl[(z_j)_{s;-i}\bigr]_{i,j=0,\ldots,n-1}
=\det\,\biggl[{1\over(z_j-s)_{-s;i}}\biggr]_{i,j=0,\ldots,n-1}
=(-1)^{n(n-1)/2}\,\biggl\{\prod_{j=0}^{n-1}(z_j)_{s;-(n-1)}\biggr\}\,\Delta_n({\bf z})\,. \eqno(3.39)$$
}

\medskip
\noindent {\bf Proof.}
Consequence of lemma 2.

\bigskip
\noindent {\bf Corollary 2.}
{\sl With $n$ a positive integer and $s$ some complex number,
$$\det\,\biggl[{1\over(z_j)_{s;-i}}\biggr]_{i,j=0,\ldots,n-1}
=\det\,\bigl[(z_j)_{s;i}\bigr]_{i,j=0,\ldots,n-1}=\Delta_n({\bf z})\,. \eqno(3.40)$$
}

\medskip
\noindent {\bf Proof.}
Consequence of lemma 1.

\bigskip
\noindent {\bf Corollary 3.}
{\sl
With $n$ a positive integer, $a$, $b$ and $s$ some complex numbers and $az_j+b\ne s,2s,\ldots,(n-1)s$,
$$\eqalignno{
\det\,\biggl[{(az_j+b)_{s;-i}\over(z_j)_{s;-i}}\biggr]_{i,j=0,\ldots,n-1}
&=\det\,\biggl[{(z_j-s)_{-s;i}\over(az_j+b-s)_{-s;i}}\biggr]_{i,j=0,\ldots,n-1} \cr
&=\biggl\{\prod_{j=0}^{n-1}
{(az_j+b)_{s;-(n-1)}\over\bigl(b+s+(n-j)(a-1)s\bigr)_{s;-j}}\biggr\}\,\Delta_n({\bf z})\,. &(3.41)\cr
}$$
}

\medskip
\noindent {\bf Proof.}
Consequence of lemma 3.

\medskip
\noindent {\bf Remarks:}

\noindent (i) It should be noted that equations (3.39), (3.40) and `almost' (3.41)
can be obtained, respectively, from equations (3.16), (3.2) and (3.25)
changing for all $w$ and $i$, $1/(w)_{s;i}$ into $(w)_{s;-i}$, although these quantities are not equal.

\noindent (ii) A proof following the same steps as proof 2 of lemma 1, and using the same linear combination
of rows ${\cal R}_i-(M_{i,0}/M_{i-1,0})\,{\cal R}_{i-1}$, can also be given for corollaries 1--3.

\noindent (iii) The extensions of lemma 1, corresponding to equations (3.3) and (3.6),
apply as well to lemmas 2 and 3 and to corollaries 1--3, see equation (C.7), e.g., with
$$\Pi_i(z):=\sum_{k=0}^{n-1} c_{i,k}\,(z)_{s;-k}\quad i=0,\ldots,n-1\qquad
\lambda:=\det\bigl[c_{i,k}\bigr]_{i,k=0,\ldots,n-1}\ne0 \eqno(3.42)$$
then
$$\det\bigl[\Pi_i(z_j)\bigr]_{i,j=0,\ldots,n-1}
=\lambda\,\det\bigl[(z_j)_{s;-k}\bigr]_{j,k=0,\ldots,n-1} \eqno(3.43)$$
and also, with $t$ some complex number,
$$\det\bigl[(z_j)_{s;t-i}\bigr]_{i,j=0,\ldots,n-1}
=\prod_{j=0}^{n-1}(z_j)_{s;t}\,\det\bigl[(z_j+t)_{s;-i}\bigr]_{i,j=0,\ldots,n-1}\,. \eqno(3.44)$$

\bigskip
\noindent {\bf Lemma 4.}
{\sl
With $n$ a positive integer and $s$ some complex number,
$$\det\bigl[(z_i+w_j)_{s;n-1}\bigr]_{i,j=0,\ldots,n-1}
=(-1)^{n(n-1)/2}\,{\bigl((n-1)!\bigr)^n\over\bigl(\prod_{j=0}^{n-1}j!\bigr)^2}\,
\Delta_n({\bf z})\,\Delta_n({\bf w})\,. \eqno(3.45)$$
}

\medskip
\noindent {\bf Proof.}
By the binomial formula (2.41), with
$M_{i,j}:=(z_i+w_j)_{s;n-1}=\sum_{k=0}^{n-1}{n-1\choose k}\,(z_i)_{s;k}\,(w_j)_{s;n-1-k}$,
the matrix $M$ reads as the product of two
matrices. Since the determinant of the product is the product of the determinants, one gets
$$\det\bigl[(z_i+w_j)_{s;n-1}\bigr]_{i,j=0,\ldots,n-1}
=\det\bigl[{n-1\choose k}(z_i)_{s;k}\bigr]_{i,k=0,\ldots,n-1}\,
\det\bigl[(w_j)_{s;n-1-k}\bigr]_{j,k=0,\ldots,n-1}\,. \eqno(3.46)$$
Taking the binomial coefficients out of the first determinant and rearranging the rows of the last determinant,
equation (3.45) follows from lemma 1.

\medskip
Note that in all lemmas and corollaries above, the determinants considered are anti-symmetric polynomials or
rational fractions of the $n$ variables $z_0,\ldots,z_{n-1}$, therefore one expects the simplest
polynomial alternant $\Delta_n({\bf z})$ to be a factor of the result. The same argument holds
for $\Delta_n({\bf w})$ in lemma 4.

\medskip
\bigskip
\noindent{\bf 4. Determinants with gamma functions or binomial coefficients as elements}

\bigskip
\noindent Using the relations (2.12)--(2.14) between the $s$-shifted factorial and the gamma function
or the binomial coefficient, the results listed below  are immediate consequences of the formulae derived
in section 3 with $s=\pm1$.
For corollaries 4--6, a direct proof following the same steps as proof 2 of lemma 1, and using the same
linear combination of rows ${\cal R}_i-(M_{i,0}/M_{i-1,0})\,{\cal R}_{i-1}$, can also be given.
It is only sketched as an example for corollary 4.
In the special case $z_j=b+aj$, with $a$ and $b$ some complex numbers, the product of differences
$\Delta_n({\bf z})$ is given by equation (B.5).

\bigskip
\noindent {\bf Corollary 4.}
{\sl
With $n$ a positive integer,
$$\eqalignno{
\det\,\bigl[\,\Gamma(z_j+i)\bigr]_{i,j=0,\ldots,n-1}
&=\Bigl\{\prod_{j=0}^{n-1}\Gamma(z_j)\Bigr\}\,\Delta_n({\bf z})\quad z_j\ne0,-1,\ldots &(4.1)\cr
\det\,\biggl[{z_j\choose i}\biggr]_{i,j=0,\ldots,n-1}
&={1\over\prod_{j=0}^{n-1}j!}\,\Delta_n({\bf z})\,. &(4.2)\cr
}$$
}

\medskip
\noindent {\bf Proof 1.}
Consequences of lemma 1.

\medskip
\noindent {\bf Proof 2.}
With the linear combination of rows ${\cal R}_i-(M_{i,0}/M_{i-1,0})\,{\cal R}_{i-1}$,
$$M_{i,j}:=\Gamma(z_j+i)\quad\rightarrow\quad(z_j-z_0)\,\Gamma(z_j+i-1)\quad i=1,\ldots,n-1\,. \eqno(4.3)$$
Then, with $z_j\ne0,-1,\ldots$, the recurrence formula on $n$ reads
$$\eqalignno{
D_n(z_0,\ldots,z_{n-1}) &:=\det\bigl[M_{i,j}\bigr]_{i,j=0,\ldots,n-1} \cr
&\phantom{:}=\Gamma(z_0)\,\Bigl\{\prod_{j=1}^{n-1}(z_j-z_0)\Bigr\}\,D_{n-1}(z_1,\ldots,z_{n-1})\,. &(4.4)\cr
}$$
Iteration of this equation, down to $D_1(z_{n-1})=\Gamma(z_{n-1})$, ends the proof of equation (4.1).

\smallskip
In the special case $z_j:=b+j\ne0,-1,\ldots$, one recovers the result already published in [1]
\footnote{${}^{15}$}{See [1] equation (A.12).}
,
$$\det\,\bigl[\,\Gamma(b+i+j)\bigr]_{i,j=0,\ldots,n-1}=\prod_{j=0}^{n-1}j!\,\Gamma(b+j)\,. \eqno(4.5)$$

\bigskip
\noindent {\bf Corollary 5.}
{\sl
With $n$ a positive integer,
$$\det\,\biggl[{1\over\Gamma(z_j+i)}\biggr]_{i,j=0,\ldots,n-1}
={(-1)^{n(n-1)/2}\over\prod_{j=0}^{n-1}\Gamma(z_j+n-1)}\,\Delta_n({\bf z}) \eqno(4.6)$$
{\sl and for $z_j\ne 0,1,\ldots,n-2$,}
$$\det\,\biggl[{1\over{z_j\choose i}}\biggr]_{i,j=0,\ldots,n-1}
=(-1)^{n(n-1)/2}\biggl\{\prod_{j=0}^{n-1}{j!\over[z_j]_{n-1}}\biggr\}\,\Delta_n({\bf z})\,. \eqno(4.7)$$
}

\medskip
\noindent {\bf Proof.}
Consequences of lemma 2.

\smallskip
In the special case $z_j:=b+j$, one gets [13]
$$\det\,\biggl[{1\over\Gamma(b+i+j)}\biggr]_{i,j=0,\ldots,n-1}
=(-1)^{n(n-1)/2}\prod_{j=0}^{n-1}{j!\over\Gamma(b+n-1+j)}\,. \eqno(4.8)$$

\bigskip
\noindent {\bf Corollary 6.}
{\sl
With $n$ a positive integer and $b$ some complex numbers, for $z_j\ne0,-1,\ldots$,
$$\det\,\biggl[{\Gamma(z_j+i)\over\Gamma(az_j+b+i)}\biggr]_{i,j=0,\ldots,n-1}
=\biggl\{\prod_{j=0}^{n-1}{\bigl(b+(n-1-j)(1-a)\bigr)_j\,
\Gamma(z_j)\over\Gamma(az_j+b+n-1)}\biggr\}\,\Delta_n({\bf z}) \eqno(4.9)$$
{\sl and for $az_j+b\ne0,1,\ldots,n-2$,}
$$\det\,\Biggl[{{z_j\choose i}\over{az_j+b\choose i}}\Biggr]_{i,j=0,\ldots,n-1}
=\biggl\{\prod_{j=0}^{n-1}{\bigl[b-(n-1-j)(1-a)\bigr]_j\over[az_j+b]_{n-1}}\biggr\}\,\Delta_n({\bf z})\,.
\eqno(4.10)$$
}
\medskip
\noindent {\bf Proof.}
Consequences of lemma 3.

\smallskip
In the special case $z_j:=c+j$, $a=1$ and $d:=b+c$ with $c\ne0,-1,\ldots$, one gets [13]
$$\det\,\biggl[{\Gamma(c+i+j)\over\Gamma(d+i+j)}\biggr]_{i,j=0,\ldots,n-1}
=\prod_{j=0}^{n-1}j!\,(d-c)_j\,{\Gamma(c+j)\over\Gamma(d+n-1+j)} \eqno(4.11)$$
where $\prod_{j=0}^{n-1}(d-c)_j=\prod_{j=0}^{n-1}(d-c+j)^{n-1-j}$.

\bigskip
\noindent {\bf Corollary 7.}
{\sl
With $n$ a positive integer and $z_j\ne0,-1,\ldots$,
$$\det\,\bigl[\Gamma(z_j-i)\bigr]_{i,j=0,\ldots,n-1}
=(-1)^{n(n-1)/2}\,\Bigl\{\prod_{j=0}^{n-1}\Gamma(z_j-n+1)\Bigr\}\,\Delta_n({\bf z})\,. \eqno(4.12)$$
}

\medskip
\noindent {\bf Proof.}
Consequence of corollary 1.

\bigskip
\noindent {\bf Corollary 8.}
{\sl
With $n$ a positive integer,
$$\det\,\biggl[{1\over\Gamma(z_j-i)}\biggr]_{i,j=0,\ldots,n-1}
={1\over\prod_{j=0}^{n-1}\Gamma(z_j)}\,\Delta_n({\bf z})\,. \eqno(4.13)$$
}

\medskip
\noindent {\bf Proof.}
Consequence of corollary 2.

\bigskip
\noindent {\bf Corollary 9.}
{\sl
With $n$ a positive integer and $az_j+b\ne n-1,n-2,\ldots$,
$$\det\,\biggl[{\Gamma(az_j+b-i)\over\Gamma(z_j-i)}\biggr]_{i,j=0,\ldots,n-1}
=\biggl\{\prod_{j=0}^{n-1}{\Gamma(az_j+b-n+1)\over\bigl(b+1+(n-j)(a-1)\bigr)_{-j}\Gamma(z_j)}\biggr\}\,
\Delta_n({\bf z})\,. \eqno(4.14)$$
}

\medskip
\noindent {\bf Proof.}
Consequence of corollary 3.

\smallskip
In the special case $z_j:=c+j$ and $a=1$ with $d:=b+c\ne n-1,n-2,\ldots$, one gets
$$\det\,\biggl[{\Gamma(d+j-i)\over\Gamma(c+j-i)}\biggr]_{i,j=0,\ldots,n-1}
=\biggl\{\prod_{j=0}^{n-1}j!\,[d-c]_j\,{\Gamma(d-n+1+j)\over\Gamma(c+j)}\biggr\}\,. \eqno(4.15)$$

\bigskip
\noindent {\bf Corollary 10.}
{\sl
With $n$ a positive integer,
$$\eqalignno{
\det\,\biggl[{\Gamma(z_i+w_j+n-1)\over\Gamma(z_i+w_j)}\biggr]_{i,j=0,\ldots,n-1}
& ={(-1)^{n(n-1)/2}\,\bigl((n-1)!\bigr)^n\over\bigl(\prod_{j=0}^{n-1}j!\bigr)^2}\,
\Delta_n({\bf z})\,\Delta_n({\bf w}) &(4.16)\cr
\noalign{\medskip}
\det\,\biggl[{z_i+w_j\choose n-1}\biggr]_{i,j=0,\ldots,n-1}
& ={(-1)^{n(n-1)/2}\over\bigl(\prod_{j=0}^{n-1}j!\bigr)^2}\,\Delta_n({\bf z})\,\Delta_n({\bf w})\,. &(4.17)\cr
}$$
}

\medskip
\noindent {\bf Proof.}
Consequences of lemma 4.

\medskip
\bigskip
\noindent{\bf 5. Some examples of applications}

\bigskip
\noindent Let us sketch some examples of applications which motivated this work, i.e.
the calculation of the probability density of the determinant (PDD) of random matrices.
Three ensembles of $n\times n$ random matrices, with $n=1,2,\ldots$, have been extensively investigated,
namely the orthogonal ($\beta=1$), unitary ($\beta=2$) and symplectic ($\beta=4$) ensembles of, respectively,
real symmetric, complex Hermitian and real quaternion self-dual matrices [14].
Then, the probability density of the eigenvalues ${\bf x}:=\{x_j\ {\rm real}\ \in{\cal D},j=0,\ldots,n-1\}$ reads
$${\rm P}_{n,\beta}({\bf x})=C_{n,\beta}\,\bigl\vert\Delta_n({\bf x})\bigr\vert^\beta
\,\prod_{j=0}^{n-1}w(x_j) \eqno(5.1)$$
where $C_{n,\beta}$ is the normalization constant, $\Delta_n({\bf x})$ is defined by equation (B.1)
and $w(x)$ is a non-negative weight function.
Quantities one computes in random matrix theory are often expressed in terms of determinants (or Pfaffians).
This is the case for the expectation value of any factorized function of the
eigenvalues, $\Phi({\bf x}):=\prod_{j=0}^{n-1}\varphi(x_j)$ [1,3].
Let us show here this result only in the simplest case $\beta=2$, namely with $d\mu(x):=w(x)\,\varphi(x)\,dx$,
one has
$$\eqalignno{
\langle\Phi\rangle &:=\int_{\cal D}d\mu(x_0)\cdots\int_{\cal D}d\mu(x_{n-1})
\,\bigl\vert\Delta_n({\bf x})\bigr\vert^2=n!\,\det\bigl[\Phi_{j,k}\bigr]_{j,k=0,\ldots,n-1} &(5.2)\cr
\Phi_{j,k} &:=\int_{\cal D}d\mu(x)\,P_j(x)\,Q_k(x) &(5.3)\cr
}$$
where $P_j$ (resp. $Q_k$) is any monic polynomial (i.e. the coefficient of its highest power is one)
of degree $j$ (resp. $k$).
Indeed, from equation (C.3), each of the two factors $\Delta_n({\bf x})$ can be expressed as
a polynomial alternant and expanded as
$\sum_{\rho\in{\cal S}_n\{0,\ldots,n-1\}}\varepsilon(\rho)\,\prod_{j=0}^{n-1} P_{\rho_j}(x_j)$,
where $\varepsilon(\rho)$ is the signature of the permutation $\rho:=\{\rho_0,\ldots,\rho_{n-1}\}$.
Thereby, one gets
$$\langle\Phi\rangle=\sum_{\rho,\sigma \in{\cal S}_n\{0,\ldots,n-1\}}\varepsilon(\rho)\,\varepsilon(\sigma)
\,\prod_{j=0}^{n-1} \int_{\cal D}d\mu(x)P_{\rho_j}(x)\,Q_{\sigma_j}(x)
=n!\,\sum_{\rho\in{\cal S}_n\{0,\ldots,n-1\}}\varepsilon(\rho)\,\prod_{j=0}^{n-1}\Phi_{\rho_j,j} \eqno(5.4)$$
completing the proof of equation (5.2).
According to the measure $d\mu(x)$ considered, one may take advantage of the freedom of choice of the monic
polynomials in order to simplify the calculations.
Thus, it may be useful to choose the set of orthogonal (or skew orthogonal for $\beta=1$ or $4$) polynomials
with respect to the weight $w(x)$ [14, 15]
\footnote{${}^{16}$}{See, e.g., [15] appendix A.14.}
.
For example, taking $\Phi$ as the identity operator, the result above with $\varphi(x)=1$ provides a
convenient way to compute the normalization constant, e.g., for $\beta=2$
$$\bigl(C_{n,2}\bigr)^{-1}=n!\,\prod_{j=0}^{n-1}\nu_j
\qquad \nu_j:=\int_{\cal D}dx\,w(x)\,P_j(x)^2 \eqno(5.5)$$
where $P_j$ are the orthogonal monic polynomials for the weight $w(x)$.

\smallskip
The calculation of the PDD,
$$g_{n,\beta}(y):=\int_{\cal D}dx_0\cdots\int_{\cal D}dx_{n-1}
\,{\rm P}_{n,\beta}({\bf x})\,\delta(y-x_0\cdots x_{n-1}) \eqno(5.6)$$
of the random matrices we consider, is based on the use of the Mellin transform.
Since this transformation explores a function only on the real non-negative half-axis,
one needs to compute the Mellin transform of the restriction to $y\ge0$ of both the even and odd parts of the PDD,
$g_{n,\beta}^\pm(y):={1\over 2}\,(g_{n,\beta}(y)\,\pm\,g_{n,\beta}(-y))$.
From equations (5.6), with $s$ some complex number, the Mellin transform of $g_{n,\beta}^\pm(y)$ reads
$$\eqalignno{
{\cal M}_{n,\beta}^\pm(s)
&:=\int_0^\infty dy\,y^{s-1}\,g_{n,\beta}^\pm(y)
={1\over2}\,\int_{\cal D}dx_0\cdots\int_{\cal D}dx_{n-1}
\,{\rm P}_{n,\beta}({\bf x})\,\prod_{j=0}^{n-1}\varphi_{\beta,s}^\pm(x) &(5.7)\cr
\varphi_{\beta,s}^\pm(x)
&:=\varepsilon^\pm(x)\,\vert x\vert^{s-1}
\qquad\varepsilon^+(x):=1\qquad\varepsilon^-(x) :={\rm sign}(x) &(5.8)\cr
}$$
namely, an expression of the type given by equations (5.2) and (5.3) when $\beta=2$, thus
$$\eqalignno{
{\cal M}_{n,2}^\pm(s) &\phantom{:}={1\over2}\,C_{n,2}\,n!
\,\det\bigl[\Phi_{j,k}^\pm(s)\bigr]_{j,k=0,\ldots,n-1} &(5.9)\cr
\Phi_{j,k}^\pm(s) &:=\int_{\cal D}dx\,w(x)\,\varphi_{2,s}^\pm(x)\,P_j(x)\,Q_k(x)\,. &(5.10)\cr
}$$

Now, one can consider several ensembles of random matrices associated with the classical orthogonal
polynomials characterized by the weight function $w(x)$ and the domain ${\cal D}$ [14]
\footnote{${}^{17}$}{See, e.g., [14] section 19.3.}
.

\noindent (i) For the frequently used Gaussian unitary ensemble [1] associated with the Hermite polynomials,
$w(x)=\exp(-x^2)$ with ${\cal D}={\cal R}$.
Choosing the polynomials $P_j$ (resp. $Q_k$) to be the monomial $x^j$ (resp. $x^k$), one finds [4]
\footnote{${}^{18}$}{See, e.g., [4] 6.1.1.}
$$\eqalignno{
\Phi_{j,k}^\pm(s)
&=\int_{-\infty}^\infty dx\,{\rm e}^{-x^2}\,\varepsilon^\pm(x)\,\vert x\vert^{s-1}\,x^{j+k}
\quad{\rm Re}\ s>0 \cr
&={1\over2}\,\bigl(1\pm(-1)^{j+k}\bigr)\,\Gamma\bigl({\textstyle{s+j+k\over2}}\bigr)\,. &(5.11)\cr
}$$
Then, the alternate elements of $\det\bigl[\Phi_{j,k}^\pm(s)\bigr]_{j,k=0,\ldots,n-1}$ being zero,
we can rearrange its rows and columns so as to collect the zero elements separate from the nonzero elements.
Note that this checkerboard structure of the determinant is true for any $w(x)\varphi(x)$ with a
well-defined parity and a domain ${\cal D}$ symmetrical with respect to $x=0$.
Thus,
$$\eqalignno{
\det\bigl[\Phi_{j,k}^+(s)\bigr]_{j,k=0,\ldots,n-1}
&=\det\bigl[\Phi_{2j,2k}^+(s)\bigr]_{j,k=0,\ldots,[(n-1)/2]}
\,\det\bigl[\Phi_{2j+1,2k+1}^\pm(s)\bigr]_{j,k=0,\ldots,[(n-2)/2]} &(5.12)\cr
\det\bigl[\Phi_{j,k}^-(s)\bigr]_{j,k=0,\ldots,n-1}
&=\cases{
(-1)^{n/2}\,\Big(\det\bigl[\Phi_{2j,2k+1}^-(s)\bigr]_{j,k=0,\ldots,n/2}\Bigr)^2 &$n$ even\cr
\noalign{\medskip}
0 &$n$ odd\cr
} &(5.13)\cr
}$$
where $[x]$ denotes the largest integer less than or equal to $x$.
From equation (5.11), the three determinants above are of the type considered in corollary 4, equation (4.5),
e.g.,
$$\eqalignno{
\det\bigl[\Phi_{2j,2k}^+(s)\bigr]_{j,k=0,\ldots,[(n-1)/2]}
&=\det\bigl[\Gamma\bigl({\textstyle{s\over2}+j+k}\bigr)\bigr]_{j,k=0,\ldots,[(n-1)/2]}
=\prod_{j=0}^{[(n-1)/2]}j!\,\Gamma\bigl({\textstyle{s\over2}+j}\bigr)\,. &(5.14)\cr
}$$

\noindent (ii) For the so-called Laguerre unitary ensemble [3], $w(x)=x^\alpha\,\exp(-x)$
with $\alpha>-1$ and ${\cal D}=[0,\infty[$.
Still choosing the polynomials $P_j$ (resp. $Q_k$) to be the monomial $x^j$ (resp. $x^k$), one finds [4]
\footnote{${}^{19}$}{See, e.g., [4] 6.1.1.}
$$\Phi_{j,k}^\pm(s)=\int_0^\infty dx\,x^\alpha\,{\rm e}^{-x}\,\vert x\vert^{s-1}\,x^{j+k}
=\Gamma(s+\alpha+j+k)\quad{\rm Re}\ s>0 \eqno(5.15)$$
Then, again with corollary 4 equation (4.5), one obtains
$$\det\bigl[\Phi_{j,k}^\pm(s)\bigr]_{j,k=0,\ldots,n-1}
=\prod_{j=0}^{n-1}j!\,\Gamma(s+\alpha+j) \eqno(5.16)$$
the result being the same for $\pm$ since the spectrum is non-negative.

\noindent (iii) For the so-called Gegenbauer unitary ensemble [3]: $w(x)=(1-x^2)^{\lambda-1/2}$
with $\lambda>{1\over2}$ and ${\cal D}=[-1,1]$.
Note that the special case $\lambda={1\over2}$ corresponds to the the so-called Legendre ensemble
with $w(x)=1$.
Still choosing the polynomials $P_j$ (resp. $Q_k$) to be the monomial $x^j$ (resp. $x^k$), one finds [4]
\footnote{${}^{20}$}{See, e.g., [4] 6.2.1 and 6.2.2.}
$$\eqalignno{
\Phi_{j,k}^\pm(s)
&=\int_{-1}^1dx\,(1-x^2)^{\lambda-1/2}\,\varepsilon^\pm(x)\,\vert x\vert^{s-1}\,x^{j+k}
\quad{\rm Re}\ s>0 \cr
&={1\over2}\,\bigl(1\pm(-1)^{j+k}\bigr)\,\Gamma\bigl({\textstyle\lambda+{1\over2}}\bigr)
\,{\Gamma\bigl({\textstyle{s+j+k\over2}}\bigr)
\over\Gamma\bigl({\textstyle\lambda+{s+j+k+1\over2}}\bigr)}\,. &(5.17)\cr
}$$
Therefore, the equations (5.12) and (5.13) are still satisfied and
the three determinants which occur are of the type considered in corollary 6 equation (4.11),
e.g.,
$$\eqalignno{
\det\bigl[\Phi_{2j,2k}^+(s)\bigr]_{j,k=0,\ldots,[(n-1)/2]}
&=\Gamma\bigl({\textstyle\lambda+{1\over2}}\bigr)^{[(n-1)/2]+1}
\,\det\biggl[{\Gamma\bigl({\textstyle{s\over2}+j+k}\bigr)\over
\Gamma\bigl({\textstyle{s+1\over2}+\lambda+j+k}\bigr)}\biggr]_{j,k=0,\ldots,[(n-1)/2]} \cr
&=\prod_{j=0}^{[(n-1)/2]}{j!\,\Gamma\bigl(\lambda+{\textstyle{1\over2}}+j\bigr)
\,\Gamma\bigl({\textstyle{s\over2}+j}\bigr)
\over\Gamma\bigl({\textstyle{s+1\over2}+\lambda+[(n-1)/2]+j}\bigr)}\,. &(5.18)\cr
}$$

\noindent (iv) For the so-called Jacobi unitary ensemble [3], $w(x)=(1-x)^a\,(1+x)^b$ with $a>1$, $b>1$
and ${\cal D}=[-1,1]$.
For $a=b=\lambda-{1\over2}$, this ensemble is identical to the Gegenbauer ensemble above.
For $a\ne b$, the problem is more complicated, in particular due to the fact that $w(x)$ is no longer an even
function.
To illustrate the use of the formulae we derived, let us calculate only the normalization constant $C_{n,2}$.
Choosing the monic polynomials $P_j(x)=(x-1)^j$ and $Q_k(x)=(1+x)^k$, one finds from equation (5.3)
with $\varphi(x)=1$
$$\Phi_{j,k}^\pm(s)=\int_{-1}^1dx\,(1-x)^{a+j}\,(1+x)^{b+k}
=(-1)^j\,2^{a+b+1+j+k}{\Gamma(a+1+j)\,\Gamma(b+1+k)\over\Gamma(a+b+2+j+k)}\,. \eqno(5.19)$$
Then, the determinant in equation (5.2) is of the type considered in corollary 5 equation (4.8),
$$\eqalignno{
\bigl(C_{n,2}\bigr)^{-1}
&=n!\,\Biggl(\prod_{j=0}^{n-1}(-1)^j\,2^{a+b+1+2j}
\,\Gamma(a+1+j)\,\Gamma(b+1+j)\Biggr)\,\det\biggl[{1\over\Gamma(a+b+2+j+k)}\biggr]_{j,k=0,\ldots,n-1} \cr
&=n!\,2^{n(n-1)+(a+b+1)n}
\,\prod_{j=0}^{n-1}{j!\,\Gamma(a+1+j)\,\Gamma(b+1+j)\over\Gamma(a+b+n+1+j)} \,. &(5.20)\cr
}$$
This result can be checked either from equation (5.5) using the constants associated with the Jacobi
polynomials [16]
\footnote{${}^{21}$}{See, e.g., [16] taking $h_j$ from equation 10.8(4) and $k_j$ from equation 10.8(5),
then $\nu_j=h_j/k_j^2\,$.}
, or from the Selberg integral [14]
\footnote{${}^{22}$}{See, e.g., [14] section 17.6.}
.

Finally, for all these unitary ensembles (except, possibly, for the currently unknown Jacobi ensemble with
$a\ne b$), the Mellin transform ${\cal M}_{n,2}^\pm(s)$ appears to be a product, or a ratio
of products, of gamma functions whose arguments are linear in $s$.
Then, from the inverse Mellin transform, the PDD is expressed in terms of Meijer G-functions [17]
\footnote{${}^{23}$}{See, e.g., [17] section 7.3 (43).}
.
For the orthogonal and symplectic ensembles the expressions are more complicated [1--3],
but we are still led to consider similar determinants.
Note that, as a by-product, one gets also the non-negative integer moments of the PDD for $q=0,1,\ldots$,
$$M_{n,\beta}(q):=\int_{\cal D} dy\,g_{n,\beta}(y)\,y^q
=\bigl(1+(-1)^q\bigr)\,{\cal M}_{n,\beta}^+(q+1)+\bigl(1-(-1)^q\bigr)
\,{\cal M}_{n,\beta}^-(q+1)\,. \eqno(5.21)$$

\smallskip
In connection with quantum coherent states, Dr K A Penson brought our attention on the boson normal
ordering problem, see [18--20] and references therein.
Let $a$ and $a\dag$ be the boson annihilation and creation operators respectively, satisfying $[a,a\dag]=1$.
The normal ordering of powers of boson monomials $\bigl((a\dag)^r\,a^s\bigr)^n$, with $n$, $r$, $s$ ($r\ge s$)
some non-negative integers involves integer sequences of numbers which are generalizations of the usual
Stirling numbers of the second kind, equation (2.9), and Bell numbers, whose values they assume for $r=s=1$,
$$\bigl((a\dag)^r\,a^s\bigr)^n:=(a\dag)^{n(r-s)}\,\sum_{k=s}^{ns}S_{r,s}(n,k)\,(a\dag)^k\,a^k
\qquad B_{r,s}(n):=\sum_{k=s}^{ns}S_{r,s}(n,k)\,. \eqno(5.22)$$
A complete theory of these sequences of numbers has been worked out.
In particular, the $B_{r,s}(n)$ can be expressed as a sum of of an infinite series of shifted factorials
(generalized Dobi\'nski formula) and moreover, can be considered as the n-th moments of a positive weight
function $W_{r,s}(x)$ with $x\ge0$,
$$B_{r,s}(n)=\int_0^\infty dx\,x^n\,W_{r,s}(x)\,. \eqno(5.23)$$
Extending $n$ to complex values and using the inverse Mellin transform, one gets from above many solutions
$W_{r,s}(x)$ of the Stieltjes moment problem [19].
Generalizing this approach to the integer sequences arising from the normal ordering of exponentiated
boson monomials, as given by equation (5.22), also provides solutions to Stieltjes moment problems.
It happens that determinants of the type we evaluate are the Hankel determinants which positivity,
if it can be proved, ensures the existence of the moment problem [20].

\smallskip
Let us add that the reader can find in [21] many methods of evaluations, lists of results and a wide bibliography
on the determinant calculus.
Beyond the evaluation of particular determinants, we want to point out that the properties of
the $s$-shifted factorials given in section 2 emphasize similarities and connections which exist with
the power function (see another example in appendix A), thereby providing compact formulae
and possibly a guide to find new relations.

\medskip
\bigskip
\noindent{\bf Acknowledgments}

\bigskip
\noindent This paper was originally motivated by the evaluation of some determinants with
gamma functions as elements which occur in works done in collaboration with M L Mehta.
P Moussa drew our attention to the importance of the exponential character in $z$ of the generating
function of shifted factorials in connection with the binomial formula; we had several
stimulating discussions on the subject.
We are also grateful to both of them for critically reading the manuscript.
Finally, we thank the referees for asking us to add some examples of applications.

\medskip
\bigskip
\noindent{\bf Appendix A. Finite sum of $s$-shifted factorials of arithmetic progression}

\bigskip
\noindent For $p$ a non-negative integer and $a$, $r$ and $s$ some complex numbers,
we compute the finite sum of $s$-shifted factorials of arithmetic progression to $n$ terms,
$$z_k:=a+k\,r\qquad S_{s;p,n}(a,r):=\sum_{k=0}^{n-1}\,(z_k)_{s;p}\quad n=1,2,\ldots \eqno(A.1)$$
using the same trick as for the sum of powers of natural numbers.
By the binomial formula (2.41),
$$(z_{k+1})_{s;p+1}=(z_k+r)_{s;p+1}
=\sum_{\ell=0}^{p+1}{p+1\choose \ell}\,(z_k)_{s;\ell}\,(r)_{s;p+1-\ell}\,. \eqno(A.2)$$
Summing up both sides of this equation for $k=0,\ldots,n-1$ yields the recurrence formula on $p$, for $n$ fixed,
$$S_{s;p,n}(a,r)={1\over(p+1)r}\,\Bigl((z_n)_{s;p+1}-(z_0)_{s;p+1}
-\sum_{\ell=0}^{p-1}{p+1\choose \ell}\,S_{s;\ell,n}(a,r)\,(r)_{s;p+1-\ell}\Bigr)
\quad p,n=1,2,\ldots\,. \eqno(A.3)$$
The first two sums are independent of $s$,
$$S_{s;0,n}(a,r)=n\qquad S_{s;1,n}(a,r)=na+{n(n-1)\over2}\,r\,. \eqno(A.4)$$

When $r=s$, then $z_k-s=z_{k-1}$, and with $s$ nonzero, an explicit expression of $S_{s;p,n}(a,s)$
can be obtained directly from the generalized Pascal triangle property (2.31),
$$\eqalignno{
S_{s;p,n}(a,s)
&={1\over (p+1)s}\,\sum_{k=0}^{n-1}\bigl((z_k)_{s;p+1}-(z_{k-1})_{s;p+1}\bigr) \cr
&={1\over (p+1)s}\,\bigl((z_{n-1})_{s;p+1}-(z_{-1})_{s;p+1}\bigr) &(A.5)\cr
}$$
where $z_{-1}=a-s$.
This result can be checked by recurrence using the general equation (A.3).
Similarly, for $r=-s$ one gets
$$S_{s;p,n}(a,-s)={1\over s(p+1)}\,\bigl((z_0)_{s;p+1}-(z_n)_{s;p+1}\bigr)\,. \eqno(A.6)$$
Thus, for $a=r=s=1$ one has, respectively, for the rising and the falling factorials
$$\eqalignno{
S_{1;p,n}(1,1) &=(1)_p+\cdots+(n)_p={(n)_{p+1}\over p+1} &(A.7)\cr
S_{-1;p,n}(1,1) &=[1]_p+\cdots+[n]_p=\left\{\matrix{
n\hfill &p=0\hfill\cr
\noalign{\medskip}
[p]_p+\cdots+[n]_p=\displaystyle{[n+1]_{p+1}\over p+1} &p=1,\ldots,n\hfill\cr
\noalign{\medskip}
0\hfill &p=n+1,\ldots\,.\hfill\cr
}\right. &(A.8)\cr
}$$

Further general properties follow from equations (2.5) and (2.7):
$$\eqalignno{
S_{s;p,n}(-a,-r) &=(-1)^p\,S_{-s;p,n}(a,r) &(A.9) \cr
\noalign{\medskip}
S_{s;p,n}(a,r) &=s^p\,S_{1;p,n}({a\over s},{r\over s})\quad s\ne0\,. &(A.10) \cr
}$$

\bigskip
\noindent {\bf Appendix B. Product of differences}

\bigskip
\noindent With the notations of equation (3.1),
the {\it product of differences} $\Delta_n({\bf z})$ is defined by
$$\Delta_n({\bf z}):=\Delta_n(z_0,\ldots,z_{n-1}):=\cases{
1 &$n=1$\cr
\noalign{\medskip}
\displaystyle{\prod_{0\le i<j\le n-1}}(z_j-z_i) &$n=2,3,\ldots$\cr}\,. \eqno(B.1)$$
The following relations are immediately obtained with $a$ and $b$ some complex numbers:
$$\eqalignno{
\Delta_n(b+az_0,\ldots,b+az_{n-1}) &=a^{n(n-1)/2}\ \Delta_n({\bf z}) &(B.2)\cr
\Delta_n\Bigl({1\over z_0}\,,\ldots,{1\over z_{n-1}}\Bigr)
&={(-1)^{n(n-1)/2}\over\prod_{j=0}^{n-1}(z_j)^{n-1}}\ \Delta_n({\bf z})\quad z_j\ne0 &(B.3)\cr
\Delta_n\Bigl({z_0\over b+az_0}\,,\ldots,{z_{n-1}\over b+az_{n-1}}\Bigr)
&={b^{n(n-1)/2}\over\prod_{j=0}^{n-1}(b+az_j)^{n-1}}\ \Delta_n({\bf z})\quad b+az_j\ne0\,. &(B.4)\cr
}$$
Finally, with $a$ and $b$ some complex numbers, in the special case $z_j:=b+aj$,
the product of differences reads
$$\Delta_n(j\mapsto b+aj)=a^{n(n-1)/2}\prod_{j=0}^{n-1}j!\,. \eqno(B.5)$$

\bigskip
\noindent {\bf Appendix C. Vandermonde's determinant}

\bigskip
\noindent It is well known [5, 15]
\footnote{${}^{24}$}{See, e.g., [5] section 14.311 or [15] section 7.1.}
that the {\it Vandermonde determinant} $\det\bigl[(z_j)^i\bigr]_{i,j=0,\ldots,n-1}$
is equal to the product of differences defined by equation (B.1), namely,
$$\det\bigl[(z_j)^i\bigr]_{i,j=0,\ldots,n-1}=\Delta_n({\bf z})\,. \eqno(C.1)$$
More generally, let us consider any set of $n$ linearly
independent polynomials in $z$ each of degree less than $n$,
$${\rm p}_i(z):=\sum_{k=0}^{n-1} c_{i,k}\,z^k\quad i=0,\ldots,n-1
\qquad \lambda:=\det\bigl[c_{i,k}\bigr]_{i,k=0,\ldots,n-1}\ne0\,. \eqno(C.2)$$
Then, since the determinant of the product is the product of the determinants, one gets
for the polynomial alternant
$$\eqalignno{
\det\bigl[{\rm p}_i(z_j)\bigr]_{i,j=0,\ldots,n-1}
&=\det\bigl[c_{i,k}\bigr]_{i,k=0,\ldots,n-1}\,\det\bigl[(z_j)^k\bigr]_{j,k=0,\ldots,n-1} \cr
&=\lambda\,\Delta_n({\bf z}) \,. &(C.3)\cr
}$$
Choosing the ${\rm p}_i$'s to be monic polynomials of degree $i$ (e.g., the monomials $z^i$),
then $c_{i,k}=0$ for $k=i+1,\ldots,n-1$ and $c_{i,i}=1$, therefore $\lambda=1$ in equation (C.2).
Now, with $b_i$ some complex numbers, it follows from the binomial formula that
$(z+b_i)^i$ is an other choice of monic polynomial of degree $i$, hence
$$\det\bigl[(b_i+z_j)^i\bigr]_{i,j=0,\ldots,n-1}=\Delta_n({\bf z}) \,. \eqno(C.4)$$
When $b_i=b$, the relation above is also a direct consequence of equations (C.1) and (B.2).

\smallskip
Since $(z^i)^{-1}=(z^{-1})^i$, with $a$ and $b$ some complex numbers,
one immediately obtains from equations (C.1), (B.3) and (B.4)
$$\eqalignno{
\det\Bigl[{1\over(z_j)^i}\Bigr]_{i,j=0,\ldots,n-1}
&={(-1)^{n(n-1)/2}\over\prod_{j=0}^{n-1}(z_j)^{n-1}}\ \Delta_n({\bf z})\quad z_j\ne0 &(C.5)\cr
\noalign{\medskip}
\det\Bigl[{(z_j)^i\over(az_j+b)^i}\Bigr]_{i,j=0,\ldots,n-1}
&={b^{n(n-1)/2}\over\prod_{j=0}^{n-1}(az_j+b)^{n-1}}\ \Delta_n({\bf z})\quad az_j+b\ne0\,. &(C.6)\cr
}$$
More generally, following the same arguments as for equation (C.3), one can consider
polynomials of the monomials introduced above (or even of any function), e.g., with
$\lambda:=\det\bigl[c_{i,k}\bigr]_{i,k=0,\ldots,n-1}$ and $az_j+b$ nonzero,
$$\det\Bigl[\sum_{k=0}^{n-1} c_{i,k}\,\Bigl({z_j\over az_j+b}\Bigr)^k\Bigr]_{i,j=0,\ldots,n-1}=
\lambda\,{b^{n(n-1)/2}\over\prod_{j=0}^{n-1}(az_j+b)^{n-1}}\ \Delta_n({\bf z})\,. \eqno(C.7)$$

\bigskip
\noindent {\bf Appendix D. Other proofs of equations (3.14), (3.21) and (3.35), (3.37)}

\bigskip
\noindent These identities can be proved by recurrence on $i$.
Let us also give a proof which illustrates another way to handle shifted factorials,
namely they can be generated by repeated derivations and/or integrations, e.g.,
$$\eqalignno{
\Bigl({d\over dx}\Bigr)^j\,x^b\bigl\vert_{x=1} &=[b]_j &(D.1)\cr
\noalign{\medskip}
\int_0^y  dy_j\cdots\int_0^{y_2}dy_1\,y_1^b\bigl\vert_{y=1} &={1\over(b+1)_j}=[b]_{-j}\,. &(D.2)
}$$

\bigskip
\noindent {\it D.1. Other proof of equation (3.14)}

\smallskip
\noindent Differentiating $j$ times $(x-1)^i\,x^{b+j-1}$ in two ways (binomial formula and chain rule derivation
of a product) [13],
$$\eqalignno{
\Bigl({d\over dx}\Bigr)^j\Bigl\{(x-1)^i\,x^{b+j-1}\Bigr\}
&=\sum_{k=0}^i(-1)^{i-k}{i\choose k}\Bigl({d\over dx}\Bigr)^j x^{b+j+k-1} \cr
&=\sum_{\ell=0}^j{j\choose \ell}\Bigl\{\Bigl({d\over dx}\Bigr)^\ell(x-1)^i\Bigr\}\,
\Bigl({d\over dx}\Bigr)^{j-\ell}x^{b+j-1} &(D.3)\cr
}$$
and setting $x=1$, only the term with $\ell=i$ is nonzero. Thereby, one gets
$$\sum_{k=0}^i(-1)^{i-k}{i\choose k}[b+j+k-1]_j=[j]_i\,[b+j-1]_{j-i} \eqno(D.4)$$
where $[j]_i$, and thus the right-hand side, vanishes for $i>j$, see equation (2.10).
Since from equations (2.12) and (2.13)
$$\eqalignno{
[b+j+k-1]_j&={\Gamma(b+j)\over\Gamma(b+i)}\,[b+i-1]_{i-k}\,(b+j)_k &(D.5)\cr
[b+j-1]_{j-i} &={\Gamma(b+j)\over\Gamma(b+i)} &(D.6)\cr
}$$
one recovers equation (3.14) (with $s=1$ for simplicity).

\bigskip
\noindent {\it D.2. Other proof of equation (3.21)}

\smallskip
\noindent Assume first $i>j$. Then, as above, one gets
$$\Bigl({d\over dy}\Bigr)^{i-j-1}\Bigl\{(y-1)^i\,y^{b+i-2}\Bigr\}\bigl\vert_{y=1}
=\sum_{k=0}^i(-1)^{i-k}{i\choose k}[b+i+k-2]_{i-j-1}=0 \eqno(D.7)$$
where the last equality is due to an overall factor $y-1$ which remains after the derivation.
When $i\le j$, integrating $j-i+1$ times $(y-1)^i\,y^{b+i-2}$ in two ways and then setting $y=1$ yield
$$\eqalignno{
\int_0^y  &dy_{j-i+1}\cdots\int_0^{y_2}dy_1\,(y_1-1)^i\,y_1^{b+i-2}\bigl\vert_{y=1}
=\sum_{k=0}^i(-1)^{i-k}{i\choose k}\,{1\over(b+i+k-1)_{j-i+1}} \cr
&=\int_0^1dy_1\,(y_1-1)^i\,y_1^{b+i-2}\int_{y_1}^1dy_2\cdots\int_{y_{j-i}}^1dy_{j-i+1}
={(-1)^i\over(j-i)!}\,{\rm B}(b+i-1,j+1) &(D.8)\cr
}$$
where ${\rm B}(z,w)$ is the beta function [4]
\footnote{${}^{25}$}{See, e.g., [4] 6.2.1 and 6.2.2.}, thus
$$\sum_{k=0}^i(-1)^{i-k}{i\choose k}\,{1\over(b+i+k-1)_{j-i+1}}
=(-1)^i\,[j]_i\,{\Gamma(b+i-1)\over\Gamma(b+i+j)}\,. \eqno(D.9)$$
Now, since from equations (2.24), (2.12) and (2.13)
$$\eqalignno{
(-1)^{i-k}[b+i+k-2]_{i-j-1} &={(-1)^{i-k}\over(b+i+k-1)_{j-i+1}} &(D.10)\cr
&={\Gamma(b+2i-1)\over\Gamma(b+j)}\times{1\over\bigl(-b-2(i-1)\bigr)_{i-k}}\,{1\over(b+j)_k} &(D.11)\cr
}$$
and furthermore
$$\eqalignno{
{\Gamma(b+i-1)\over\Gamma(b+i+j)}
&={\Gamma(b+2i-1)\over\Gamma(b+j)}\times{1\over(b+i-1)_i\,(b+j)_i} &(D.12)\cr
}$$
the sum over $k$ in equation (D.7) for $i>j$ and (D.9) for $i\le j$
does correspond to the sums considered in equation (3.21) (with $s=1$ for simplicity).
Note that since $[j]_i$ vanishes for $i>j$ and with equation (D.10), the relation (D.9) is true in all cases.

\bigskip
\noindent {\it D.3. Other proof of equations (3.35), (3.37)}

\smallskip
\noindent Assume first $i>j$. With now two variables $x$ and $y$, as above, one gets
$$\eqalignno{
\Bigl({\partial\over\partial x}\Bigr)^j &\Bigl({\partial\over\partial y}\Bigr)^{i-j-1}
\Bigl\{(xy-1)^i\,x^{c+j-1}\,y^{d+i-2}\Bigr\}\bigl\vert_{x=1\atop y=1} \cr
&=\sum_{k=0}^i(-1)^{i-k}{i\choose k}[c+j+k-1]_j\,[d+i+k-2]_{i-j-1} \cr
&=\Bigl({d\over dx}\Bigr)^j\biggl\{x^{c-d}\Bigl({d\over dx}\Bigr)^{i-j-1}
\Bigl\{(x-1)^i\,x^{d+i-2}\Bigr\}\biggr\}\bigl\vert_{x=1}=0 &(D.13)\cr
}$$
where the last equality is due to an overall factor $x-1$ which remains after the derivation over $x$.
When $i\le j$, differentiating $j$ times with respect to $x$ and integrating $j-i+1$ times
over $y$ the expression $(xy-1)^i\,x^{d-c+i-2}$ in two ways and then setting $x=1$ and $y=1$ yield
$$\eqalignno{
\int_0^y &dy_{j-i+1}\cdots\int_0^{y_2}dy_1\,\Bigl({\partial\over\partial x}\Bigr)^j\,
\Bigl\{(xy_1-1)^i\,x^{c+j-1}\,y_1^{d+i-2}\Bigr\}\bigl\vert_{x=1\atop y=1} \cr
&=\sum_{k=0}^i(-1)^{i-k}{i\choose k}\,{[c+j+k-1]_j\over(d+i+k-1)_{j-i+1}} \cr
&=\sum_{\ell=0}^j{j\choose\ell}\int_0^1dy_1\,
\Bigl\{\Bigl({\partial\over\partial x}\Bigr)^\ell(xy_1-1)^i\Bigr\}
\Bigl\{\Bigl({d\over dx}\Bigr)^{j-\ell}x^{c+j-1}\Bigr\}y_1^{b+i-2}
\int_{y_1}^1dy_2\cdots\int_{y_{j-i}}^1dy_{j-i+1}\bigl\vert_{x=1} \cr
&={1\over(j-i)!}\,\sum_{\ell=0}^i(-1)^{i-\ell}{j\choose \ell}\,[i]_l\,[c+j-1]_{j-l}\,
{\rm B}(d+i+\ell-1,j-\ell+1) &(D.14)\cr
}$$
where ${\rm B}(z,w)$ is the beta function. Thereby, after some elementary algebra,
using equations (2.24), (2.12) and (2.13), one gets
$$\eqalignno{
\sum_{k=0}^i &(-1)^{i-k}{i\choose k}\,{[c+j+k-1]_j\over(d+i+k-1)_{j-i+1}} \cr
&=[j]_i\,{\Gamma(c+j)\over\Gamma(c+i)}\,{\Gamma(d+i-1)\over\Gamma(d+i+j)}\,
\sum_{\ell=0}^i(-1)^{i-\ell}{j\choose \ell}\,(d+i-1)_\ell\,[c+i-1]_{i-\ell} \cr
&={\Gamma(d+2i-1)\over\Gamma(d+j)}\,{\Gamma(c+j)\over\Gamma(c+i)}
\times{[j]_i\,(d-c)_i\over(d+j)_i\,(d+i-1)_i} &(D.15)\cr
}$$
where the last equality follows from the binomial formula (2.46). Now, since
$$\eqalignno{
[c+j+ &k-1]_j\,[d+i+k-2]_{i-j-1}={[c+j+k-1]_j\over(d+i+k-1)_{j-i+1}} &(D.16)\cr
&={\Gamma(d+2i-1)\over\Gamma(d+j)}\,{\Gamma(c+j)\over\Gamma(c+i)}\times
{[c+i-1]_{i-k}\over[d+2(i-1)]_{i-k}}\,{(c+j)_k\over(d+j)_k} &(D.17)\cr
}$$
the sum over $k$ in equation (D.13) for $i>j$ and (D.15) for $i\le j$
does correspond to the sum over $k$ in equation (3.35) (with $s=1$ for simplicity).
Note that since $[j]_i$ vanishes for $i>j$ and with equation (D.16), the relation (D.15) is true in all cases.

\medskip
\bigskip
\noindent {\bf References}

\bigskip

\item{[1]} Mehta M L and Normand J-M 1998
Probability density of the determinant of a random Hermitian matrix
{\it J. Phys. A: Math. Gen.} {\bf31} 5377-91

\item{[2]} Delannay R and Le Ca\"er G 2000
Distribution of the determinant of a random real-symmetric matrix from the Gaussian orthogonal ensemble
{\it Phys. Rev.} E {\bf62} 1526-36

\item{[3]} Normand J-M and Mehta M L 2004
Probability density of the determinant of some random matrix ensembles
{\it preprint} SPhT-T04/039

\item{[4]} Abramowitz M and Stegun I A 1972
{\it Handbook of Mathematical Functions}
(New York: Dover)

\item{[5]} Gradshteyn I S and Ryzhik I M 1980
{\it Table of Integrals, Series, and Products}
(New York: Academic)

\item{[6]} Comtet L 1970
{\it Analyse combinatoire I} and {\it II}
(Presses Universitaires de France, Paris)

\item{[7]} Rota G-C and Mullin R 1970
On the Foundations of Combinatorial Theory: III. Theory of Binomial  Enumeration,
in {\it Graph Theory and its Applications} ed B Harris
(New York: Academic)

\item{[8]} Roman S M and Rota G-C 1978
The Umbral Calculus
{\it Adv. Math.} {\bf27} 95-188

\item{[9]} Aigner M 1980
{\it Combinatorial Theory} Reprint of the 1979 Edition
(Berlin: Springer)

\item{[10]} Rosen K H, Michaels J G, Gross J L, Grossman J W and Shier D R 2000
{\it Handbook of Discrete and Combinatorial Mathematics}
(Boca Raton, FL: CRC)

\item{[11]} Moussa P 2003 private communication

\item{[12]} Bateman H 1953
{\it Higher Transcendental Functions} vol 1
(New York: McGraw-Hill)

\item{[13]} Mehta M L 2003 private communication

\item{[14]} See, e.g., Mehta M L 1991
{\it Random Matrices}
(New York: Academic)

\item{[15]} See, e.g., Mehta M L 1989
{\it Matrix Theory}
(Les Editions de Physique, 91944 Les Ulis Cedex, France)

\item{[16]} Bateman H 1953
{\it Higher Transcendental Functions} vol 2
(New York: McGraw-Hill)

\item{[17]} Bateman H 1954
{\it Tables of Integral Transforms} vol 1
(New York: McGraw-Hill)

\item{[18]} Blasiak P, Penson K A and Solomon A I 2003
The boson normal ordering problem and generalized Bell numbers
{\it Ann. Comb.} {\bf7} 127-39

\item{[19]} Penson K A and Solomon A I 2003
Coherent state measures and the extended Dobi\'nski relations
{\it Symmetry and Structural Properties of Condensed Matter: Proc. 7th Int. School of Theoretical Physics
(Myczkowce, Poland, Sep. 2002)}
ed T Lulek, B Lulek and A Wal (Singapore: World Scientific) p 64
({\it Preprint} quant-ph/0211061)

\item{[20]} Penson K A, Blasiak P, Duchamp G, Horzela A and Solomon A I 2004
Hierarchical Dobi\'nski-type relations via substitution and the moment problem
{\it J. Phys. A: Math. Gen.} {\bf37} 3475-87

\item{[21]} Krattenthaler C 1999
Advanced determinant calculus
{\it S\'eminaire Lotharingien Combin} {\bf 42} paper B42q 67 pp
({\it Preprint} math.CO/9902004)

\end